\documentclass[aip]{revtex4-1}

\usepackage[hidelinks]{hyperref}
\usepackage{graphicx}
\usepackage{amsmath}
\usepackage{amssymb}
\usepackage{epstopdf, epsfig}
\usepackage[dvipsnames]{xcolor}
\usepackage{float}
\usepackage{grffile}
\usepackage{soul}
\usepackage{multirow}
\usepackage[english]{babel}

\ifx\TikzNotPDF\undefined
\else
\usepackage{tikz}
\usepackage{pgfplots}
\pgfplotsset{compat=newest}
\usetikzlibrary{plotmarks}
\usepgfplotslibrary{patchplots}

\usetikzlibrary{calc}
\usepgfplotslibrary{external}
\tikzsetexternalprefix{./}

\pgfplotsset{
	x tick label style={
		/pgf/number format/1000 sep=},
}
\fi
\graphicspath{{./figure/}}

\hypersetup{
	colorlinks   = true, 
	urlcolor     = blue, 
	linkcolor    = black, 
	citecolor   = black 
}

\begin{document}
	
	\title{Data compression for turbulence databases using spatio-temporal sub-sampling and local re-simulation}
	
	\author{Zhao Wu}
	\author{Tamer A. Zaki}
	\author{Charles Meneveau}
	
	\affiliation{ 
		Department of Mechanical Engineering, Johns Hopkins University, Baltimore, MD 21218, USA
	}
	
	\date{\today}
	
	\begin{abstract}
		Motivated by specific data and accuracy requirements for building numerical databases of turbulent flows, data compression using spatio-temporal sub-sampling and local re-simulation is proposed. Numerical re-simulation experiments for decaying isotropic turbulence based on sub-sampled data are undertaken. The results and error analyses are used to establish parameter choices for sufficiently accurate sub-sampling and sub-domain re-simulation.   
	\end{abstract}
	
	\maketitle 
	
	\section{Introduction}\label{sec:intro}
	
	In the field of computational fluid dynamics, the study of turbulent flows based on data generated using Direct Numerical Simulations (DNS) has occupied a prominent place in the literature over the past several decades \citep{Kim1987,Moin1998, Livescu2008, Yeung2012, Bermejo-Moreno2013}
	DNS provides spatial and temporal resolution down to the smallest and fastest eddies of a turbulent flow. Therefore, the Reynolds number achievable by DNS is limited by computing power and memory, and has been growing roughly at the rate expected from Moore's law. The amount of data generated by DNS is growing accordingly \citep{Yeung2015,Lee2015,Lee2018, Yamamoto2018,You2019}. For instance, a simulation of turbulent flow outputting four field variables (e.g. the three velocity components and pressure) on $2000^3$ spatial grid points and integrated over, say, $5\times 10^4$ time-steps, will generate several Petabytes (PB) of data.  Researchers thus store only a few selected snapshots of the flow during the simulations, and primarily rely on run-time analysis tools that are decided prior to the computation if time resolved phenomena are to be studied. As a result, when new questions and concepts arise, massive simulations must be performed over and over. Moreover, when storing snapshot data for later analysis, the traditional means of sharing available data after DNS, e.g. \citet{DelAlamo2003}, assumes that the data are downloaded as flat files and consequently a user has to worry about formats and provide the computational resources for analysis.   
	
	As a means to address these problems that challenge further growth of DNS and accessibility of data, modern database technologies have begun to be applied to DNS-based turbulence research. For instance, the Johns Hopkins Turbulence Database (JHTDB, \url{http://turbulence.pha.jhu.edu}) \citep{Perlman2007, Li2008},  has been constructed and has been in operation for about a decade, as an open public numerical laboratory. The system hosts about 1/2 PB DNS data including 5 space-time resolved data sets and several others with a few snapshots available. Users have Web-services facilitated access to the data, among others using a ``virtual sensors'' approach in which a user specifies the position and time at which data are requested and the system returns properly interpolated field data.  Other derived quantities such as gradients \citep{Li2008} and fluid trajectories \citep{Yu2012} are also available, typically delivered to within single-precision machine accuracy. A hallmark of the system is the ability of users to access very small targeted subsets of the data without having to download the entirety of the data.  The system has been successful at democratizing access to some of the world's largest high-fidelity DNS of canonical turbulent flows. JHTDB data have been used in over 120 peer-reviewed journal articles since its inception, about 25 in 2018 alone.  
	
	In recent years, the scale of DNS data has continued to grow further. The largest simulations now generate data on about $O(10^4)$ grid points in each of the three directions, so storing multiple time steps to capture time evolution becomes very challenging, even in efficiently built databases.  For example, only one snapshot of the $Re_\tau=5200$ channel flow DNS \citep{Lee2015} is 1.8 TB,  and one snapshot of the $8192^3$ isotropic turbulence \citep{Yeung2015} is 8 TB in single precision. Hence,  storing temporally evolved fields over meaningful time horizons is becoming unfeasible. Storing even only one large-scale turnover time of the $8192^3$ isotropic turbulence data set would require storing about 80 Petabytes. Over the next several years, it can be anticipated that even larger scale DNS will be performed, generating Exabytes of data, far out of reach of anticipated facilities and the approaches on which JHTDB is currently based.  
	
	It is therefore necessary to explore innovative tools for compressing  simulation data for use in conjunction with databases. Most of the general-purpose data compression algorithms are based on analyzing the data representation, and can generally be classified as lossless or lossy. 
	Lossless data compression utilizes the statistical redundancy \citep[e.g.][]{Huffman1952,Ziv1977}, while lossy data compression is to remove unnecessary data, e.g., JPEG \citep{ISO.1993} and MP3 \citep{ISO.1994}. Lossless data compression tools are promising but for turbulence data where the flow's small-scale structures contain non-trivial information at each grid point, the compression ratios can be expected to be somewhat limited. While we continue current efforts along this direction and can expect further improvements, more aggressive tools will be required for the very large datasets envisioned in the near future. Regarding lossy compression, it is certainly appropriate for visualization and other applications where less fidelity is acceptable. However, if one wishes, e.g. to capture accurately velocity gradients, lossy compression algorithms in which the accuracy of primary variables is degraded, say, at the fourth decimal point, will already lead to significant errors in gradients and will thus be insufficient for the purposes of turbulence research.  
	
	
	It bears recalling that JHTDB enables users to receive interpolated data between spatial and temporal grid points, using polynomial functions (Lagrange, spline, Hermite). Far more aggressive data compression could be achieved if data could be stored more sparsely in both space and time. However, when a user requests localized pieces of data that fall between coarsely stored positions and/or times, one would need to revert to the dynamical equations (i.e. Navier-Stokes) to perform a physics-based rather than a polynomial based interpolation. 
	
	In this paper we explore and establish requirements for such a  data compression method, named ``Spatio-Temporal Sub-sampling and sub-domain Re-simulation'' (STSR). The method aims at enabling users to recover data at close to machine accuracy (single-precision), based on very coarsely stored data. While the method can greatly compress the amount of data to be stored, such savings have to be balanced by the additional cost of processor (CPU or GPU) expense needed later on  to accommodate user queries. 
	
	Initial efforts attempting to reproduce DNS data using local re-simulation (technical details to be provided below) have shown a surprisingly narrow and stringent range of conditions under which re-simulation in a sub-domain can generate data at the desired accuracy. That is to say, re-simulation that reproduces DNS at close to single-precision machine accuracy, the desired baseline accuracy level, is more difficult to achieve than one may expect. Any small deviations from the conditions to be developed can be shown to lead to significant errors. It will be observed that the errors do not arise due to chaotic dynamics as we do not observe exponential divergence of state-space trajectories or exponential growth of errors over time. The lack of chaotic divergence of dynamics may be due to the strong constraints introduced by boundary conditions prescribed around closed sub-domains. Instead, errors are introduced due to small details of numerical implementation, discretization, and order of operations that at first glance may appear small and trivial but that can cause rather significant differences in results.   
	
	Therefore, the present paper aims to document the technical methodologies and tests performed with considerable attention to detail. Section \ref{sec:resimulation} introduces the basic idea of data compression for 
	turbulence databases using spatio-temporal sub-sampling and local re-simulation (STSR).
	The desire to enable re-simulations over localized spatial domains precludes the use of spectral methods based on global basis functions. In this work, we explore the use of one of the most common discretization tools in CFD: second order finite differencing. The numerical scheme adopted in the present computations is described in Section \ref{sec:numerical_scheme}. The methodology is tested in the context of a well-understood and relatively simple flow: Section \ref{sec:decaying_isotropic_turbulence} describes the decaying isotropic turbulence used as a test case to document performance of several approaches and variants. In Section \ref{sec:results}, the influence of the boundary conditions on reproducibility of the simulations, up to the desired level of machine precision, is examined. The re-simulation errors  are studied in Section \ref{sec:error_analysis} in more detail, and their dependence on artificially introduced noise in  boundary conditions is established in order to better understand requirements for reaching desired levels of accuracy, which are slightly relaxed from machine accuracy down to relative errors at the order of $\sim 10^{-5}$ based on practical considerations. Section \ref{sec:summary} showcases an application using the recommended parameters.  Finally, conclusions are presented in Section \ref{sec:conclusion}. The paper is limited to an account of the findings regarding methodology and requirements in the context of a simple flow at moderate computational scale. Construction of a large turbulence database system using the proposed spatio-temporal sub-sampling and local re-simulation querying method is left as a future task. 
	
	Although this work is focused on turbulence in incompressible flows, extensions of the basic idea and methodological requirements to other fields of computational physics appear possible.
	

	\section{Sub-sampling and local re-simulation}\label{sec:resimulation}
	
\begin{figure}
		\centering
		\raisebox{4.3cm}{(a)}\includegraphics[width=5cm]{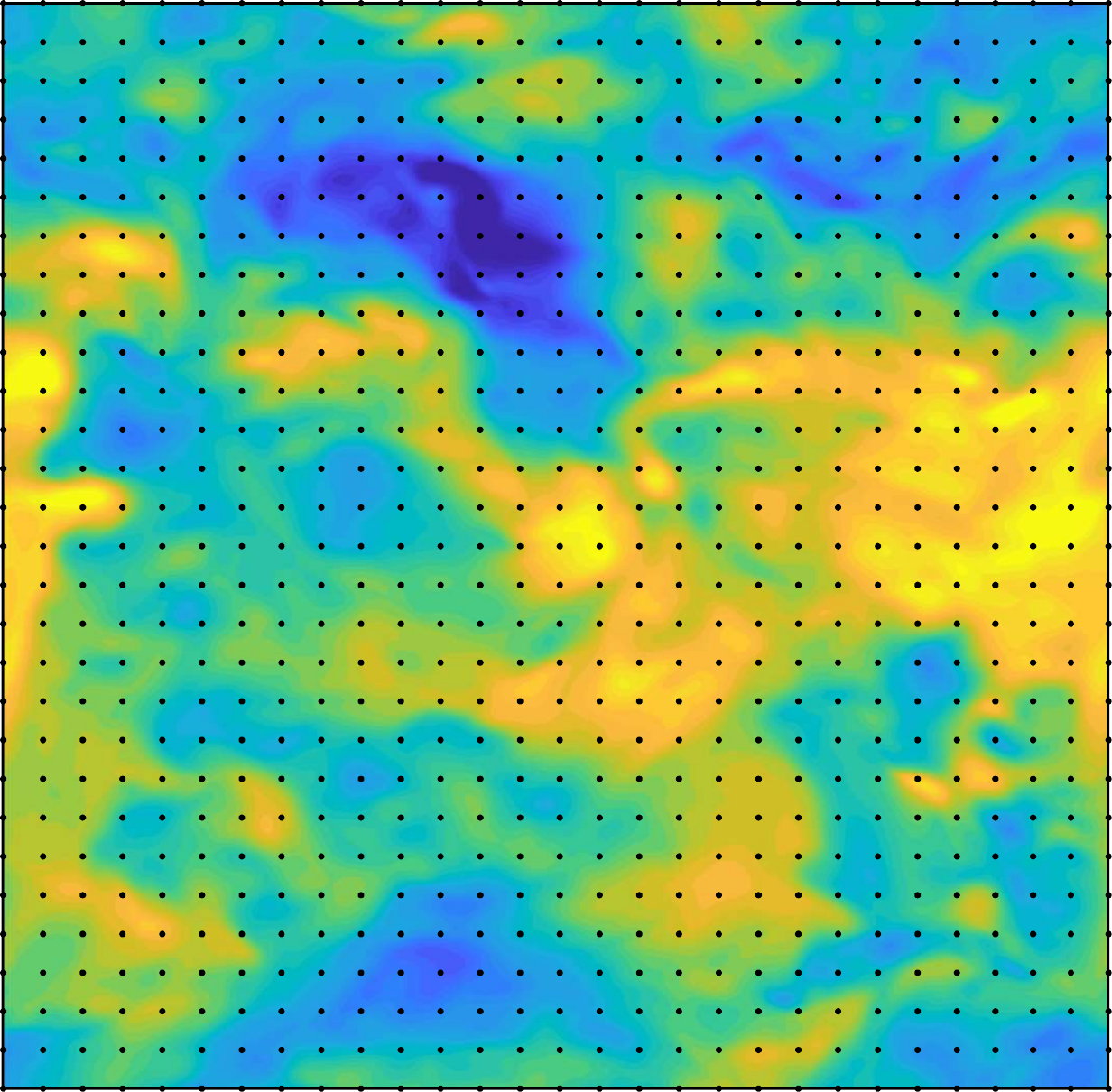}\quad
		\raisebox{4.3cm}{(b)}\includegraphics[width=5cm]{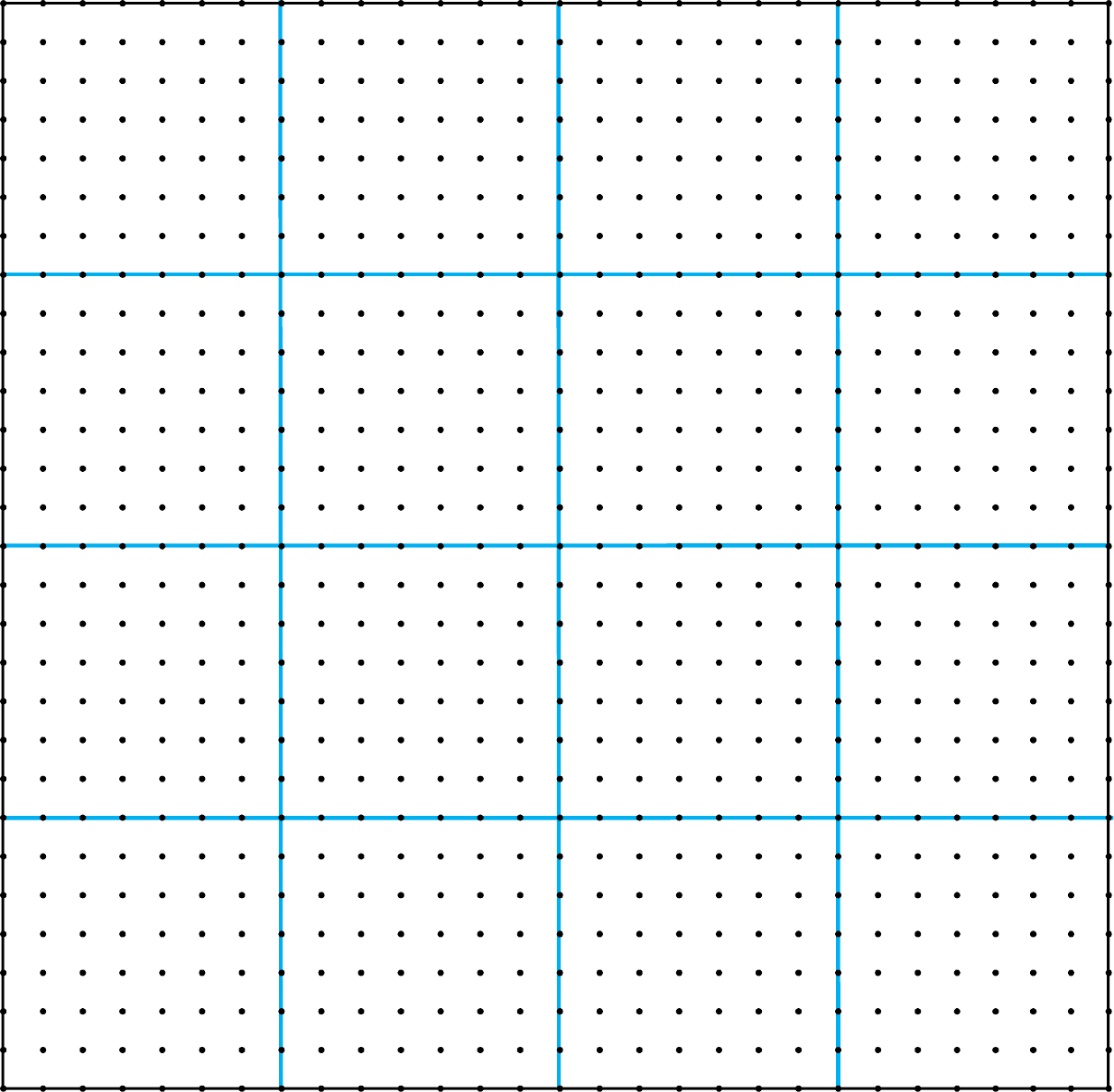}\\[10pt]
		\raisebox{1cm}{(c)}\includegraphics[width=10.5cm]{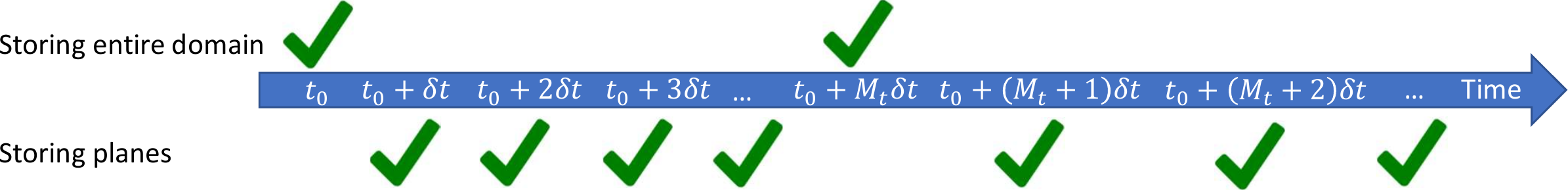}\\[10pt]
		\caption{(a) Entire DNS domain containing  a large $(N^3)$ number of grid points. (b) The entire DNS domain are divided into small cube regions by the blue lines.
			(c) The storage scheme of the  spatio-temporal subsampling for re-simulation. The data in the entire domain are stored at every $M_t$ time step. The data on the planes (blue lines) and on the outer planes (black lines), are stored at every time step. 
		}\label{fig:resim}
	\end{figure}

	In this section, the basic concept of the proposed STSR approach is explained, together with an estimate of the data compression that can be achieved. Figure \ref{fig:resim} is a two-domensional schematic of a DNS domain and the storage scheme of the data to enable later re-simulation. The flow domain inside the box in Figure \ref{fig:resim}(a) represents the entire, or global, domain of the original simulation, e.g. from a simulation of isotropic turbulence, channel flow, boundary layer, etc.
	The global domain consists of a large number of grid points; in 3-D, say, $N^3 = N_x N_y N_z$. By enforcing initial and boundary conditions on the global domain boundaries, the simulation is advanced forward in time, at a time-step $\delta t$.  The objective is to store a limited amount of data at each time-step in order to enable re-simulation of a sub-region of the global domain. For this purpose, the global domain is divided into small sub-volumes marked by the blue boundaries (figure \ref{fig:resim}(b)) corresponding to planes in a three-dimensional domain.
	For simplicity, the sub-volumes here have the same shape and dimensions but the discussion and general results to be presented can be considered quite general.
	While the main simulation is performed, the state vector (i.e. velocity and pressure fields for incompressible flow) is stored on these planes. If the size of an individual re-simulation sub-domain is $M_s$, in 3-D there will be $3(N/M_s)$ such planes, each of size $N^2$. 
	
	Moreover, in order to limit the CPU cost of re-simulation, after a number of time-steps, the state vector data are stored at every grid point in the global domain. This occurs every $M_t$ time steps, i.e. after a time equal to $M_t \, \delta t$ (see Figure \ref{fig:resim}(c)).  
	In the rest of this paper, $t_n = n \,\delta t$ represents the physical time, while $n$ represents the time step of the DNS.  For a simulation lasting a total time $T$, the total number of full 3D fields to be stored is thus equal to $\sim T/ (M_t \delta t)$. 
	
\begin{figure}
		\centering
		\includegraphics[width=5cm]{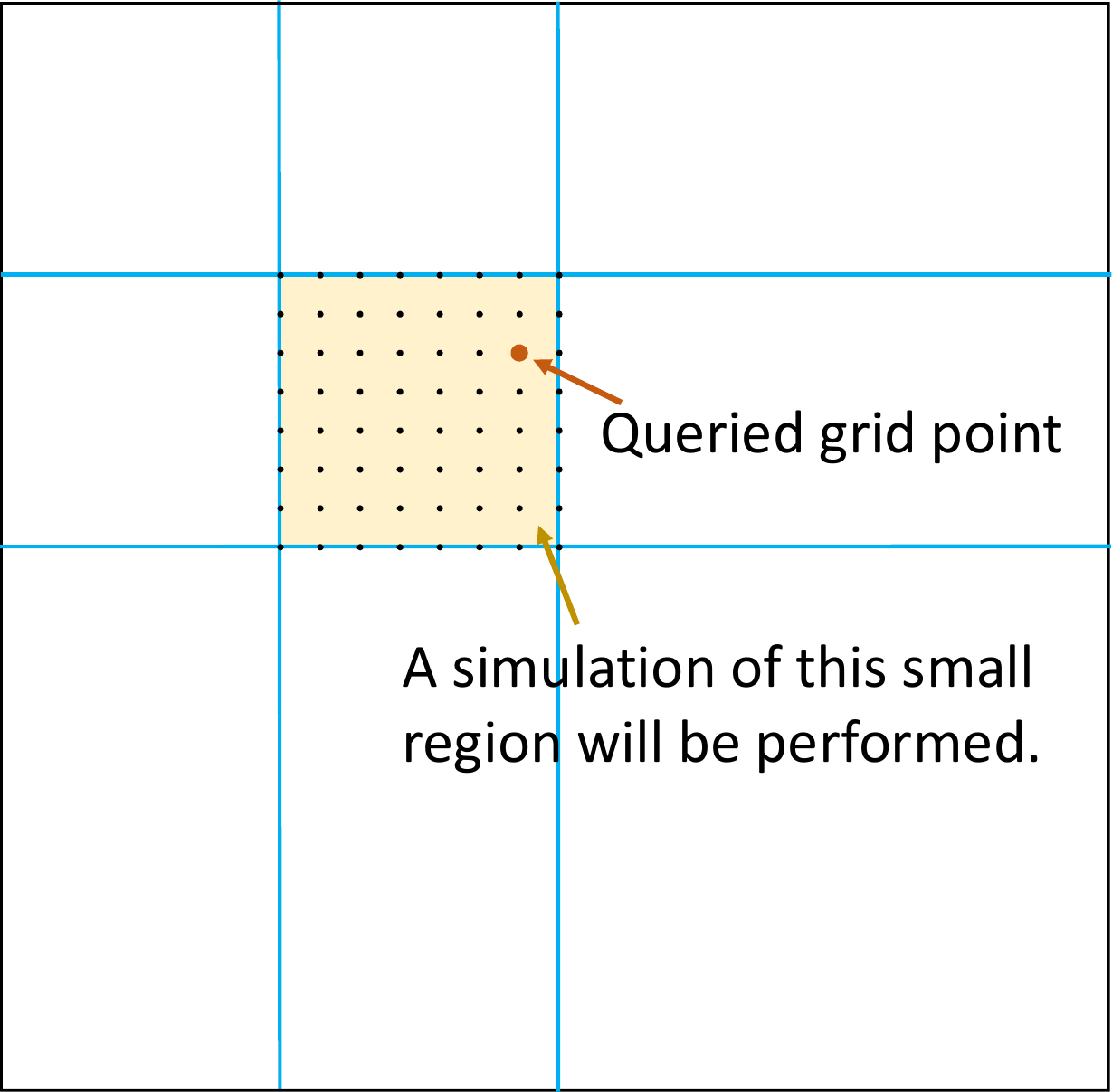}
		\caption{When data is required on gridpoints that are not stored in the database, a re-simulation of a small region which includes the queried grid point is performed to obtain the data.}\label{fig:resim2}
	\end{figure}

	After the direct simulation in the global domain has been completed and the sub-sampled data stored, data at a specific spatial and temporal location ($\boldsymbol{x}, t$) may be required, for example to examine local flow states in particularly interesting sub-regions of the flow or to track particles through the flow. In general these locations do not correspond to stored data, and the data must be evaluated by re-evaluating the flow evolution in the host sub-volume and time interval (Figure \ref{fig:resim2}).
	%
	%

	Similar to the global domain, the flow in the re-simulation sub-domain is governed by the continuity and Navier-Stokes equations. The numerical solution requires the initial and boundary conditions. Suppose there exists an integer $n$   such that $(t_0+n M_t\delta_{\rm dns} <t< t_0+(n+1)M_t\delta t_{\rm dns})$, i.e. the time at which data are sought $t$ lies between two instances where the entire global domain was stored. The data stored at $(t_0+n M_t\delta_{\rm dns})$ can then be used as the initial condition, and the plane data on the sub-domain boundary that was stored at every time step between times $(t_0+n M_t\delta_{\rm dns})$ and $t$ provide the boundary conditions needed for re-simulation. 
	Unless otherwise stated, the original simulation and its re-simulation will adopt the same time step for forward integration of the governing equations.
	
	To fix notation, in the rest of this work the ``global domain'' refers to the domain of the original simulation (the black enclosing box in figure \ref{fig:resim2}); a ``sub-domain'' refers to the much smaller region containing a queried point or sets of points (the yellow region in figure \ref{fig:resim2}); and ``re-simulation'' refers to numerical solution of the governing equation in this sub-domain using initial and boundary conditions extracted during the original computation and stored in the STSR database.
	
	With the proposed approach, only a small fraction of data is stored and the fields can be re-constructed on demand from simulations within the small sub-regions. The data compression (inverse) ratio $c$ can be estimated as 
	\begin{equation}\label{eq:compression_ratio}
	c \approx \frac{N^3+3 N^2(N/M_s)(M_t-1)}{N^3 M_t} = \frac{1}{M_t}\left(1-\frac{3}{M_s}\right)+\frac{3}{M_s}
	\end{equation}
	where $N$ is the number of grid points in each direction in the entire domain.
	
	Hence, if for example $M_s=128$ is used, and we store only every $M_t = 200$ full 3D fields, the total storage requirement is about 2.8\% of the original data. Performing the re-simulation in the $M_s^3$ sub-domain is certainly much faster than doing a re-simulation in the original full 3D volume: the CPU cost of re-simulation is approximately 
	$M_t(12M_s^3+M_s^3\log_2M_s)$.
	Depending on the ratio of cost of storage and computation, as well as depending on patterns of data queries and usage,  the optimal values of $M_s$ and $M_t$ could vary significantly. For now we simply observe that the $8192^3$ grid database with $\sim 10^4$ time steps  mentioned in the introduction requiring over 80 PB of storage, would require only about 2.2 PB if stored using sub-sampling with $M_s=128$ and $M_t=200$, and the computational cost of the re-simulation is only $\mathcal{O}(10^{-6})$ of the cost of the full simulation.
	
	The approach becomes particularly attractive in   studies where only small sub-regions of the flow need to be interrogated later on. For example, in particle tracking studies, one only needs velocities in the immediate vicinity of particles to be used for interpolation. In other studies, researchers may want to zoom into areas where extreme events such as core of vortices or high dissipation take place. Or, one may wish to obtain a one-dimensional spectrum along some representative lines through the flow requiring data only along those lines rather than the entire domain.
	In such scenarios, storing the entire data or having to perform re-simulation in the entire domain would be unnecessary and waste computational/storage resources.
	

	\section{Numerical scheme and flow configuration}\label{sec:numerical_scheme}
	
	
	Incompressible flow of a Newtonian fluid satisfies the continuity and Navier-Stokes equations written here in skew-symmetric form, 
	\begin{equation}\label{eq:cont}
	\nabla \cdot \boldsymbol{u}=0,
	\end{equation}
	\begin{equation}\label{eq:NS}
	\frac{\partial \boldsymbol{u}}{\partial t}+\frac{1}{2} (\nabla \cdot (\boldsymbol{u} \otimes \boldsymbol{u})+(\boldsymbol{u} \cdot \nabla)\boldsymbol{u})=- \nabla p +\nu \nabla^2 \boldsymbol{u},
	\end{equation}
	where $\boldsymbol{u}=(u,v,w)^T$ is the velocity vector, $t$ is time, and $\nu$ is the fluid kinematic viscosity. The three velocity components $u$, $v$ and $w$ correspond to the $x$, $y$ and $z$ directions, respectively, and $p$ is pressure divided by density. The advection term in (\ref{eq:NS}) is expressed in the skew-symmetric form which conserves kinetic energy and reduces aliasing errors \citep{Kravchenko1997}. However, other forms of the advection term can also be adopted.
	
	\subsection{Temporal and spatial discretization}
	
	We adopt a fractional-step algorithm that decouples the velocity and pressure, which is widely used in solving incompressible flow problems. The $\delta p$-form is used \citep{VanKan1986, Bell1989, Nicolaou2015} where the intermediate velocity $\boldsymbol{u}^*$, which is not necessarily divergence free, is computed with the pressure gradient term included in the prediction step:
	\begin{align}\label{eq:prediction}
		\frac{\boldsymbol{u}^*-\boldsymbol{u}^{(n-1)}}{\delta t} =& -[\alpha_c C(\boldsymbol{u}^{(n-1)})+\beta_c C(\boldsymbol{u}^{(n-2)})] \nonumber \\
		& + \nu [\alpha_v L(\boldsymbol{u}^{(n-1)})+ \beta_v L(\boldsymbol{u}^{(n-2)})+ \gamma_v L(\boldsymbol{u}^{*})] \\
		& - G(p^{(n-1)}). \nonumber
	\end{align}
	In the above equation, $\delta t$ is the size of the time step and superscript $(\cdot)^{n}$ denotes the $n$-th step.  The term $C(\boldsymbol{u})$ is the discretized advection operator, $L$ is the discretized Laplacian operator and $G$ is the discretized gradient operator. The constant coefficients $(\alpha, \beta, \gamma)$ depend on the temporal discretization.  For example, the advection term can be advanced in time explicitly using explicit Euler ($\alpha_c=1,\  \beta_c=0$) or second-order Adams-Bashforth (AB2) scheme ($\alpha_c=3/2,\ \beta_c=-1/2$); the viscous term can be advanced using Euler ($\alpha_v=1,\ \beta_v=0,\ \gamma_v=0$), AB2 ($\alpha_v=3/2,\ \beta_v=-1/2,\ \gamma_v=0$) or implicit Crank-Nicolson (CN) scheme ($\alpha_v=1/2,\ \beta_v=0,\ \gamma_v=1/2$).
	
	The intermediate velocity $\boldsymbol{u}^*$ is then projected onto a divergence-free field
	\begin{equation}\label{eq:correction}
	\boldsymbol{u}^{(n)}=\boldsymbol{u}^* - \delta t \, G\phi^{(n)},
	\end{equation}
	where $\phi$ is the solution of a Poisson equation 
	\begin{equation}\label{eq:poisson}
	DG\phi^{(n)}=\frac{D \boldsymbol{u}^*}{\delta t},
	\end{equation}
	and $D$ is the discretized divergence operator.
	Finally, the pressure is updated using, 
	\begin{equation}\label{eq:pres}
	p^{(n)}=p^{(n-1)}+\phi^{(n)}.
	\end{equation}
	As seen, $\phi^{(n)}$ is the difference of the pressure at two consecutive time steps, thus the above algorithm is called $\delta p$ form.
	A variant of the projection method referred to as the $p$-form \citep{Chorin1968, Kim1985} ignores the pressure gradient term in the prediction step (\ref{eq:prediction}), and therefore $\phi^{(n)}$ in the Poisson equation (\ref{eq:poisson}) is an approximation of the full pressure at the new time step, i.e.\,$p^{(n)}=\phi^{(n)}$. 
	
	An notable difference between the herein adopted $\delta p$ and the $p$ forms is in the boundary conditions:
	(i) the boundary condition of the elliptic pressure equation is the pressure difference in the $\delta p$-form, and the pressure in the $p$-form;
	(ii) in terms of the velocity, in order to ensure second-order accuracy, one should enforce $\boldsymbol{u}^*=\boldsymbol{u}_{\Gamma}$ on the boundary of the computational domain $\Gamma$ in the $\delta p$-form, but $\boldsymbol{u}^*=\boldsymbol{u}_{\Gamma}+\delta t \, G p_\Gamma^{(n-1)}$ in the $p$-form.
	In the present study, the $\delta p$-form is adopted throughout. Although not presented here, use of the $p$-form does not affect our results nor conclusions.

\begin{figure}
		\centering
		\includegraphics[width=7cm]{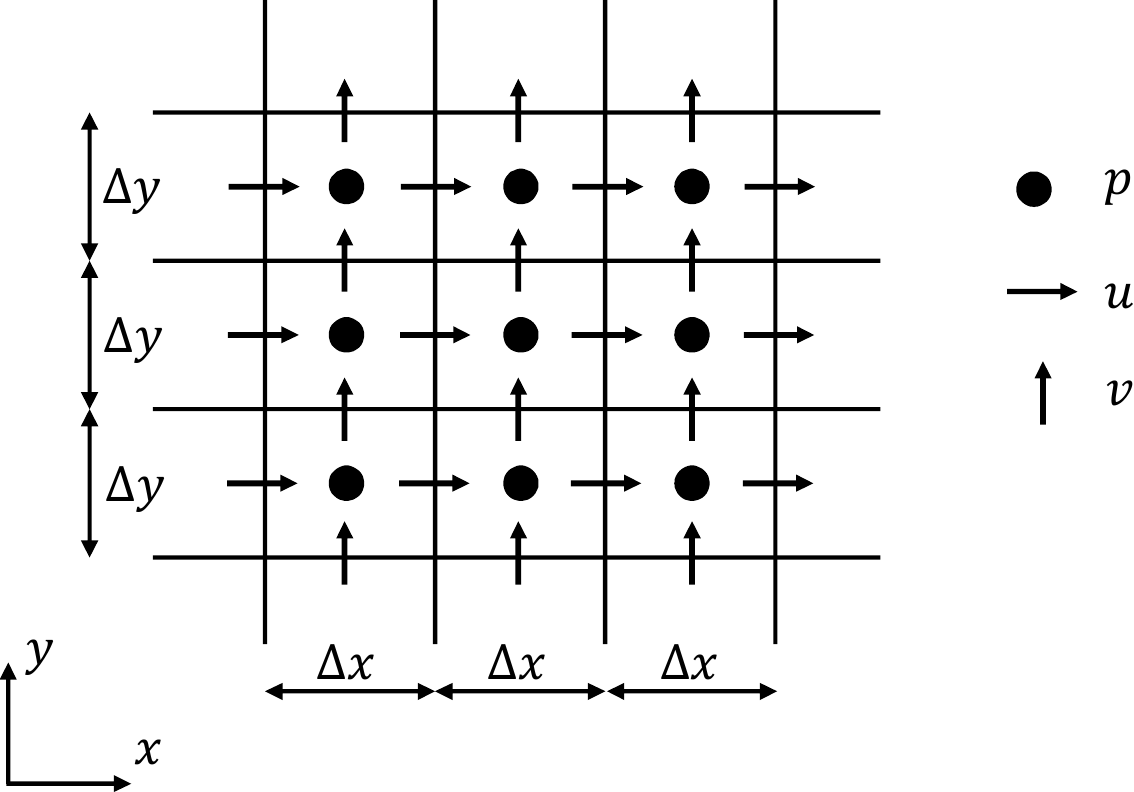}\\
		\caption{Sketch of staggered mesh in two dimensions}
		\label{fig:staggered_grid}
	\end{figure}

	
	A staggered grid \citep{Harlow1965} is used in order to avoid checkerboard pressure oscillations\textemdash see figure \ref{fig:staggered_grid}. The spatial derivatives are approximated with second-order central finite differences. The three components of the momentum equations are evaluated at different cell faces, while the divergence-free condition is satisfied at the cell centers, where the pressure is located.
	
	The velocity boundary conditions are applied on the cell faces, and can be either periodic, Dirichlet or Neumann type. \citet{Gresho1987} and \citet{Abdallah1988} showed that the Poisson equation (\ref{eq:poisson}) with Dirichlet or Neumann boundary conditions has the same solution, to within a constant.  As such, the boundary condition for $\phi$ could also be periodic, Dirichlet or Neumann and is applied at the cell faces. Unless otherwise stated, the Neumann boundary condition is adopted in the Poisson equation in the re-simulations.
	
	In light of the computational cost of the pressure equation (\ref{eq:poisson}), it is important to ensure that the re-simulation does not compromise any of the efficiency of the global solver. For instance, if the global domain is triply periodic, Fourier transform can be adopted in all three dimensions and the solution of (\ref{eq:poisson}) is inexpensive.  The re-simulation sub-domain is, however, not periodic; we nonetheless adopt a fast Poisson solver using discrete sine and cosine transforms \citep{Schumann1988}. Details on the pressure Poisson solver used in re-simulations are provided in Appendix A.
	
	\subsection{Flow configuration: decaying isotropic turbulence}\label{sec:decaying_isotropic_turbulence}
	
	
	The flow adopted in this work as an example application of STSR is decaying isotropic turbulence in three dimensions. The global domain has dimensions $2\pi \times 2\pi \times 2\pi$, and is discretized uniformly using $256^3$ grid points ($N=256$); the grid spacing is $h=\Delta x =0.02454$. The domain is periodic in all three spatial directions. Time integration of the viscous and convective terms starts with one Euler step at the initial condition, and is subsequently evolved using AB2. A snapshot from an $1024^3$ isotropic turbulence data set (\url{https://doi.org/10.7281/T1KK98XB}) in the Johns Hopkins Turbulence Database is used as the initial condition, subsampled every 4 grid points. After a transient of a few hundred time steps, the entire velocity and pressure fields are stored and designated as the initial condition ($t=0$, $n=0$) of our set of numerical experiments.  
	
\begin{figure}
		\centering
		\raisebox{3.6cm}{(a)}\includegraphics[height=3.7cm]{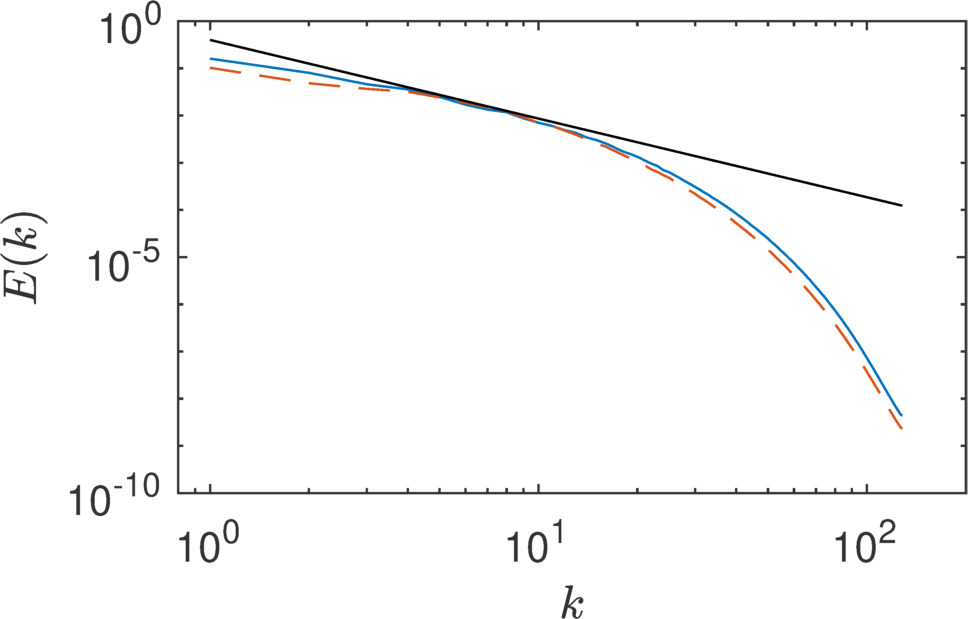}~
		\raisebox{3.6cm}{(b)}\includegraphics[height=3.7cm]{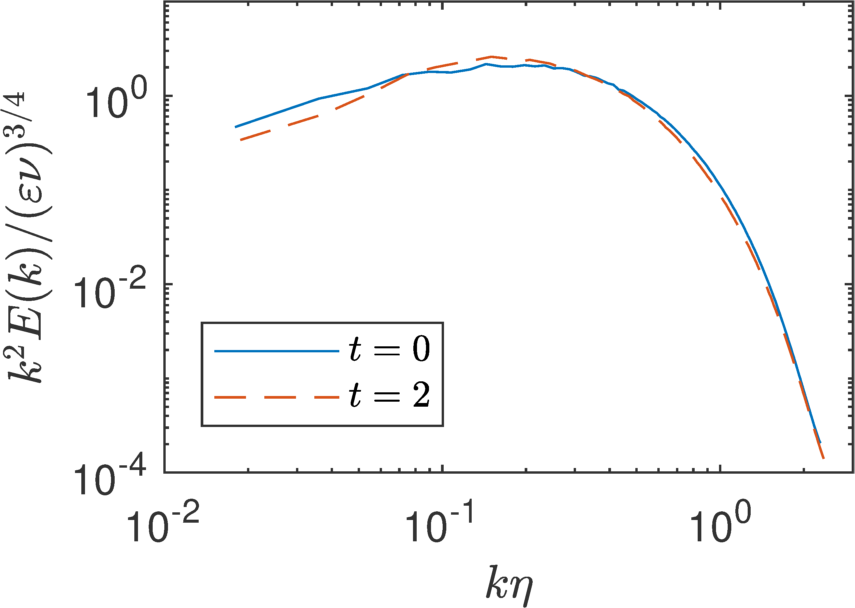}\\
		\caption{Radial (a) kinetic energy and (b) dissipation spectra at the start of the simulation $t=0$ and the end of the simulation $t=2$. The black straight line in (a) has a slope of -5/3.
		}
		\label{fig:spectra}
	\end{figure}

	The kinematic viscosity is set to $\nu=2\times10^{-3}$ in order to provide appropriate resolution of the viscous scale at the initial time. 
	Five different time steps will be used, $\delta t = \{4, 2, 1, 0.5, 0.25\}\times10^{-3}$. 
	Simulations are advanced from $t=0$ to $t=2$. 
	The root-mean-square velocity, dissipation, Taylor-scale based Reynolds number and Kolmogorov scales at the initial and final times of the simulation of decaying isotropic turbulence are listed in Table \ref{tab:statistics}. These  were verified to be accurate to within four digits for the various choices of the time step; the reported CFL values are  based on the largest $\delta t= 4 \times 10^{-3}$. The kinetic energy and dissipation spectra at the start and the end of the simulation, $t=\{0,2\}$, are shown in figure \ref{fig:spectra}. The dissipation spectra are displayed in Kolmogorov units, showing that the simulation is very well resolved in space (note that the spatial resolution is much better than than in the JHTDB original data even if using less points since here we simulate a much lower Reynolds number with a much higher $\nu$).
	
	{
		\setlength{\tabcolsep}{3.5pt}
		\begin{table}
			\begin{center}
				\begin{tabular}{ccccccc}
					\hline
					{Time} & {RMS vel.}
					& {Dissipation} & Re-number & {Kolm. scale} & \multicolumn{2}{c}{CFL} \\[5pt]
					\hline
					$t$& $u'$& $\varepsilon$ & $R_\lambda$ & $\eta$& $u'\delta t/\Delta x$ & $u_{\max}\delta t/\Delta x$ \\[5pt]
					\hline
					$0$ & 0.6024 & 0.0770 & 113.24 & 0.01795 & 0.0982 & 0.4013 \\[7pt]
					$2$ & 0.5185 & 0.0645 & 91.67 & 0.01876 & 0.0845 & 0.3699 \\[7pt]
					\hline
				\end{tabular}
				\caption{Statistics of decaying isotropic turbulence in the global domain ($256^3$). The statistics are the same to within four digits for the four different time steps used, except for the quoted CFL numbers which are based on the case $\delta t=4\times10^{-3}$.}
				\label{tab:statistics}
			\end{center}
		\end{table}
	}
	
	While performing the simulation in the global domain, data are stored at every time step to be used for later analysis and comparison with re-simulation results.
	For the sample re-simulations and numerical experiments to be described in the next section, a sub-domain consisting of $32^3$ grid points is selected (i.e. $M_s=32$) located at a random location within the global domain. To compare results from re-simulation to the original global domain simulation, a normalized local error is defined according to
	\begin{equation}\label{eq:error}
	\epsilon_\varphi(x,y,z,t)=\frac{|\varphi_{os}(x,y,z,t)-\varphi_{rs}(x,y,z,t)|}{\text{rms}(\varphi_{os})},
	\end{equation}
	where $\varphi$ could be $u$, $v$, $w$ or $p$, $\text{rms}(\cdot)$ is the root-mean-square (r.m.s) value within the sub-domain, ``$os$'' refers to the original simulation, and ``$rs$'' refers to the re-simulation.
	We focus on the $L^\infty$ errors evaluated as function of time within the sub-domain, $\epsilon_{\varphi,\infty}(t)$, which is a stringent upper bound on the re-simulation errors.
	
	\section{Preliminary results}\label{sec:results}
	
	As a first test we consider re-simulation in a $32^3$ sub-domain starting from the initial condition at $t=0$. 
	One can use the velocity and pressure fields at $n=0$ as the re-simulation initial condition. The simplest is to use velocities stored at every successive time steps on the six bounding planes as the velocity boundary condition. For pressure, we use the  gradient of \textit{pressure differences} (not pressure\textemdash see equation \ref{eq:poisson}) on the sub-domain boundaries $\Gamma$ for $n>0$ as the re-simulation boundary conditions. Specifically, these boundary conditions at time step $n$ are $\boldsymbol{u}^{(n)}_{rs,\Gamma} = \boldsymbol{u}^{(n)}_{os,\Gamma}$ for velocity and $\left(\partial \phi^{(n)}_{rs}/\partial \boldsymbol{n}\right)_\Gamma = \left(\partial \phi^{(n)}_{os}/\partial \boldsymbol{n}\right)_\Gamma$ for pressure increment. Above, $\boldsymbol{n}$ denotes the outward pointing normal unit vector to the boundary $\Gamma$ (distinct from time step $n$). 
	
\begin{figure}
		\centering
		\raisebox{4.4cm}{(a)}\includegraphics[height=4.8cm]{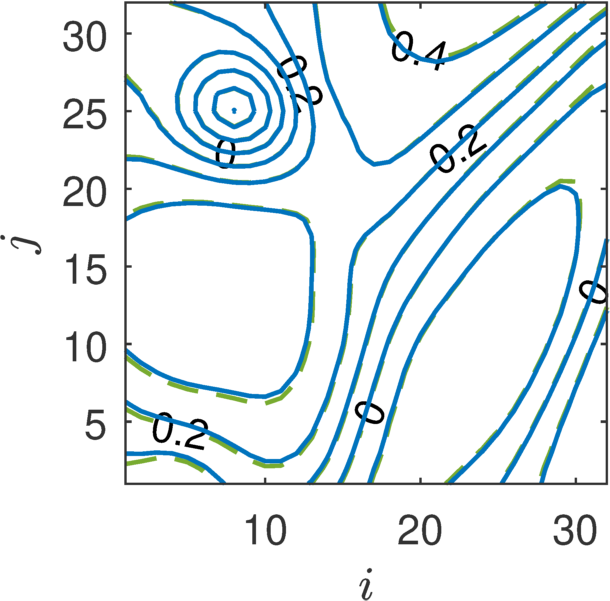}~~
		\includegraphics[height=4.8cm]{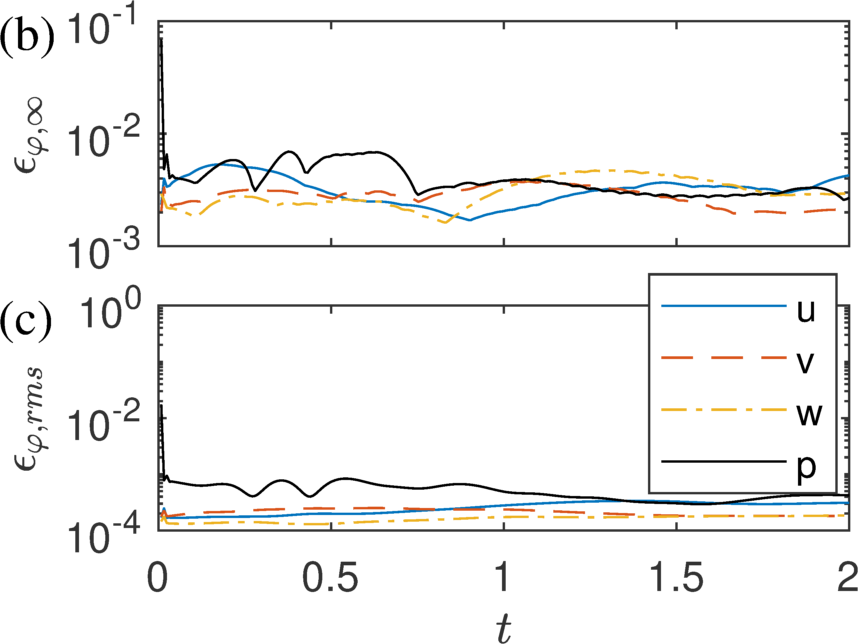}\\
		\caption{(a) Contour plot of  pressure distribution on a randomly selected slice in the $32^3$ sub-domain re-simulation at a randomly selected, representative, time step. The dash contour lines are the original simulation, while the solid contour lines are the re-simulation. (b)$\varepsilon_{\varphi,\infty}$ as function of time  $t$. (c) r.m.s error $\varepsilon_{\varphi,rms}$ as function of time  $t$.  In the re-simulation, the velocity boundary conditions are $\boldsymbol{u}$ and the pressure boundary condition is Neumann type. All plots are for the case $\delta t=4\times10^{-3}$. 
		}
		\label{fig:err_u_vs_time}
	\end{figure}

	Using these initial and boundary conditions, the re-simulation is integrated in time between $t=0$ all the way to $t=2$ (i.e. for 500 time-steps for the case $\delta t = 4 \times 10^{-3})$. The results are compared with the original simulation (see figure \ref{fig:err_u_vs_time} for the $\delta t=4\times10^{-3}$ case).
	Figure \ref{fig:err_u_vs_time}(a) shows a comparison of the pressure distribution on a representative plane and time. While overall the agreement may appear good, there are some noticeable differences especially near the lower left and upper right boundaries. 
	
	More quantitatively, the maximum error ($L^\infty$) and r.m.s error over the sub-domain are shown as functions of time in Figures \ref{fig:err_u_vs_time}(b) and (c).  The error is large already at the first re-simulation time-step and then remains of similar order of magnitude. Both the $L^\infty$ and r.m.s errors are of order $10^{-3}$, which is too large compared to our stated desired level of accuracy. 
	
	An interesting observation is that the errors do not grow exponentially, suggesting that the errors are not caused by chaotic dynamics as one may have initially suspected based on the non-linear character of the governing equations. Recall that here the boundary conditions are imposed at all times which may prevent exponential divergence between the original and re-simulation dynamics.  We have experimented with a number of parameters such as the time step and spatial resolution, and the basic conclusion remains that the errors are significant and far from the desired accuracy for our database application. Aiming to reduce these errors, we analyze the source of the discrepancy and identify the appropriate choice of implementing initial and boundary conditions in order to greatly reduce these errors.
	
	\subsection{Re-simulation boundary conditions: \textbf{\textit{u}} versus \textbf{\textit{u}}$^*$}
	\label{sec:u_or_ustar}
	
\begin{figure}
		\centering
		\includegraphics[width=\textwidth]{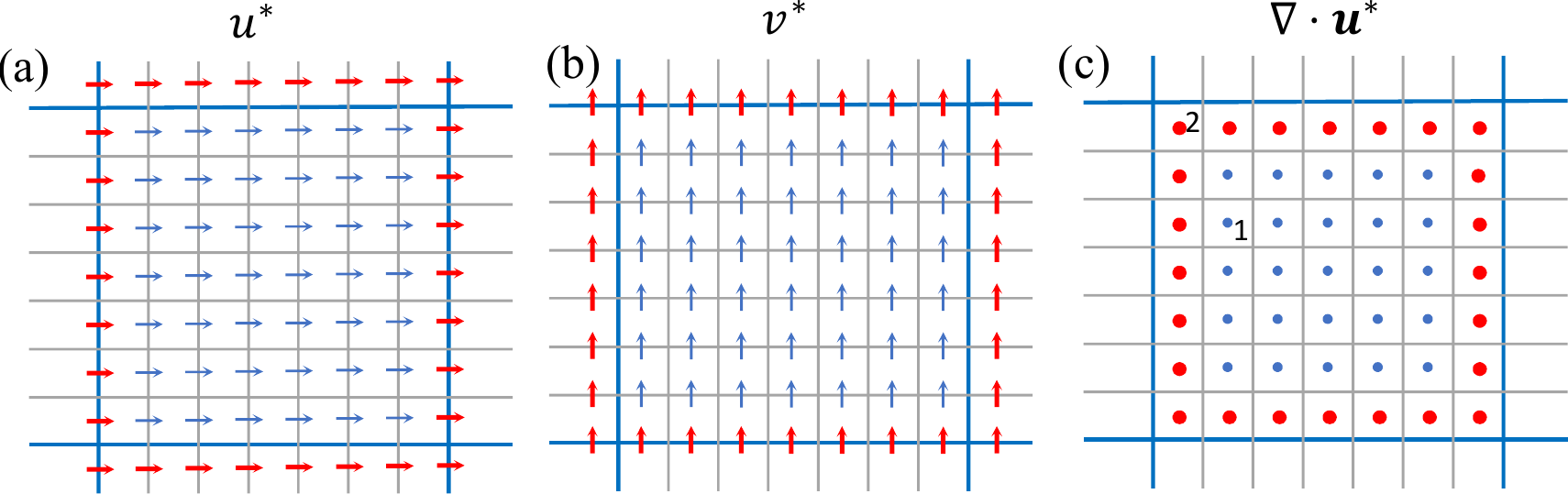}\\
		\caption{Comparisons of (a) $u^*$, (b) $v^*$ and (c) $\nabla \cdot \boldsymbol{u}^*$ between the re-simulation and the original simulation. Blue (thin) and red (thick) symbols denote the quantities in re-simulation match/mismatch to the original simulation data.
		}
		\label{fig:u_ana}
	\end{figure}

	
	Consider the re-simulation procedure from the initial condition $n=0$ to the first time step $n=1$.  
	At time step $n=0$, the initial conditions are based on $\boldsymbol{u}_{os}^{(0)}$ and $p_{os}^{(0)}$ of the global computation, and therefore  the re-simulation matches that state exactly.
	%
	Since $\boldsymbol{u}^{(0)}_{rs}$ and $p^{(0)}_{rs}$ match the global simulation, the convective, diffusive and pressure gradient terms \textit{inside} the re-simulation sub-domain are correct.
	%
	Because the momentum equations are both integrated with an Euler method in the original simulation and the re-simulation to the first time step $n=1$, $\boldsymbol{u}^*$ \textit{inside} the sub-domain is the same as in the original simulation (blue thin arrows in figure \ref{fig:u_ana}(a,b)). 
	Meanwhile, $\boldsymbol{u}^{(1)}_{os}$ on $\Gamma$ are applied as the velocity boundary conditions. However, the data on and outside the sub-domain boundary are also $\boldsymbol{u}^*$ in the original simulation, since they lie within the global domain. Thus, the re-simulation does not match the original computation on and outside the sub-domain boundary (red thick arrows in figure \ref{fig:u_ana}(a,b)).
	%
	The source term of the Poisson equation is then computed, and the comparison with the original simulation is shown in figure \ref{fig:u_ana}(c). Considering two grid points as examples, the source term at point 1 is calculated from surrounding values of $\boldsymbol{u}^*_{rs}$, all of which are identical to the original simulation.  Thus the source term is correct (blue small dots). However, at point 2, the values of ${u}^*$ at left and $v^*$ above are different from the original simulation, thus the source term at this grid point differs from the global solver (red big dots).
	%
	The Poisson equation with perturbed source term is solved and $\boldsymbol{u}_{rs}$ and $\phi_{rs}$ therefore contain errors.
	
	The above discussion shows that the choice of velocity boundary conditions leads to errors in the re-simulation outcome, as reported in figure \ref{fig:err_u_vs_time}. The remedy is to adopt $\boldsymbol{u}^*_{os}$ as the velocity boundary condition in the re-simulation procedure.  
	
\begin{figure}
		\centering
		\raisebox{4cm}{(a)}\includegraphics[width=5.5cm]{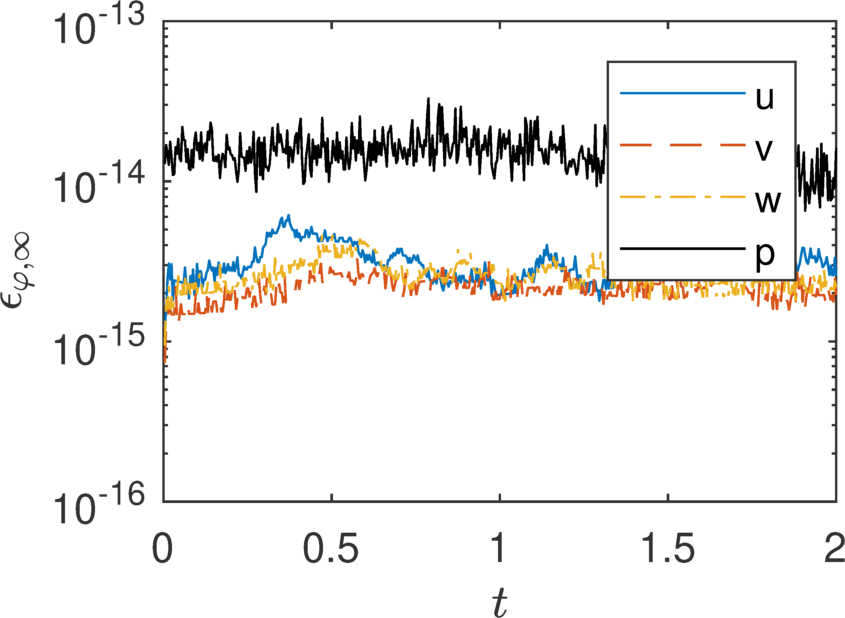}~
		\raisebox{4cm}{(a)}\includegraphics[width=5.5cm]{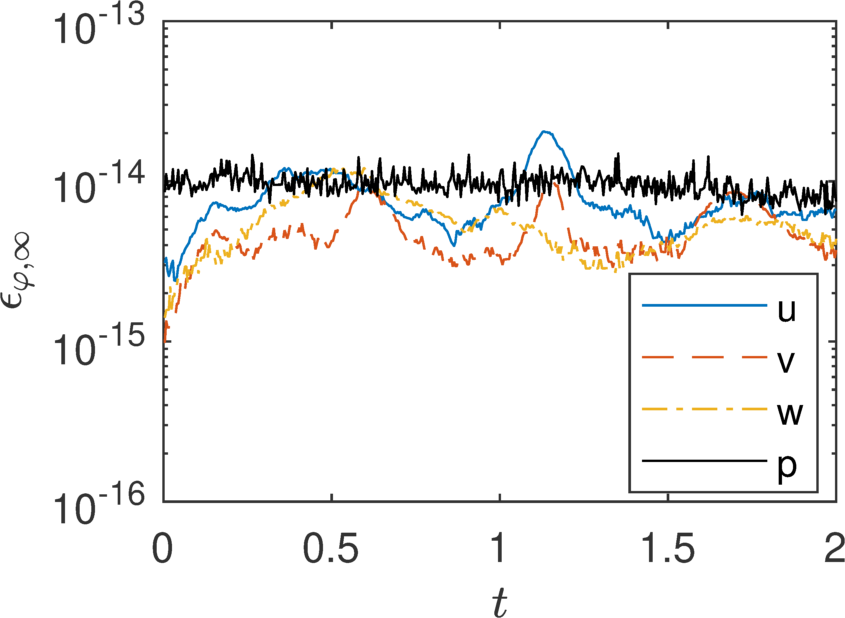}\\
		\caption{Errors $\varepsilon_{\varphi,\infty}$ as function of time (for the case $\delta t=4\times10^{-3}$). In the re-simulation, the velocity boundary conditions are $\boldsymbol{u}^*$, and the pressure boundary condition is (a) Neumann type and (b) Dirichlet type.}
		\label{fig:err_ustar_vs_time}
	\end{figure}

	Thus switching procedure, now the values of $\boldsymbol{u}^*_{os}$ at the boundaries of the sub-domains were stored during the global simulation.  These were subsequently used for boundary conditions in the local re-simulation procedure.  The resulting $L^\infty$ errors are reported in figure \ref{fig:err_ustar_vs_time}(a). Indeed the re-simulation velocities and pressure  agree with the global computation results exactly, to within machine precision.
	
	\citet{Gresho1987} and \citet{Abdallah1988} showed that Dirichlet and Neumann pressure boundary conditions are equivalent, to within a constant. We confirmed the same behavior for the re-simulations by performing a test with pressure Dirichlet boundary conditions $\phi_{rs}=\phi_{os}$ on the sub-domain boundary $\Gamma$. The re-simulation errors, shown in figure \ref{fig:err_ustar_vs_time}(b), are still at machine accuracy, the same as those in the re-simulations with the pressure Neumann boundary conditions (note that in both cases $\boldsymbol{u}^*_{rs}= \boldsymbol{u}^*_{os}$ is enforced on $\Gamma$).
	
	
	\subsection{Crank-Nicolson scheme}\label{sec:CN_scheme}
	
	In simulations of non-homogeneous flows such as wall-bounded turbulent flows, the viscous term may limit the time step  due to the stability restriction. Therefore, this term is often discretized in time using Crank-Nicolson (CN) scheme, in order to mitigate the stability restriction. Using CN, equation (\ref{eq:prediction}) is approximated with the alternating direction implicit (ADI) method according to, 
	\begin{equation}\label{eq:prediction_CN}
	(1-A_x)(1-A_y)(1-A_z)\boldsymbol{u}^* =\delta t[-Conv.+\frac{1}{2}\nu L(\boldsymbol{u}^{(n-1)})- G(p^{(n-1)})]+\boldsymbol{u}^{(n-1)},
	\end{equation}
	where $A_x=\frac{1}{2} \nu \delta t L_x$, $A_y=\frac{1}{2} \nu \delta t L_y$, $A_z=\frac{1}{2} \nu \delta t L_z$, $Conv.=\alpha_c C(\boldsymbol{u}^{(n-1)})+\beta_c C(\boldsymbol{u}^{(n-2)})$ is the integrated advection term, and $L_x$, $L_y$ and $L_z$ are the discretized Laplacian operators in the $x$, $y$ and $z$ directions. 
	The procedure for solving the above equation consists of evaluating $\boldsymbol{u}^{*}$ in each of the three directions successively: 
	(i) solve for $\boldsymbol{u}^{*1}$ in the $x$ direction, where $(1-A_x)\boldsymbol{u}^{*1} =$right hand side of equation (\ref{eq:prediction_CN}) with $x$ boundary conditions; 
	(ii) solve for $\boldsymbol{u}^{*2}$ in the $y$ direction, where $(1-A_y)\boldsymbol{u}^{*2} =\boldsymbol{u}^{*1}$ with $y$ boundary conditions;
	(iii) solve for $\boldsymbol{u}^{*}=\boldsymbol{u}^{*3}$ in the $z$ direction, where $(1-A_z)\boldsymbol{u}^{*3} =\boldsymbol{u}^{*2}$ with $z$ boundary conditions. 
	In Section \ref{sec:u_or_ustar}, it was demonstrated that $\boldsymbol{u}^{*}$ should be the velocity boundary condition if both the original and re-simulation algorithms are explicit Euler/AB2.  When CN/ADI is adopted however, different intermediate velocity boundary conditions are required. Specifically, $\boldsymbol{u}^{*1}$ should be applied on the boundaries during the inversion of the $x$-diffusion term, $\boldsymbol{u}^{*2}$ should be applied on the boundaries during the solution in the $y$ direction, and $\boldsymbol{u}^{*}=\boldsymbol{u}^{*3}$ should be applied on the boundaries in the final $z$ direction.
	
\begin{figure}
		\centering
		\raisebox{2cm}{(a)}\includegraphics[width=3.3cm]{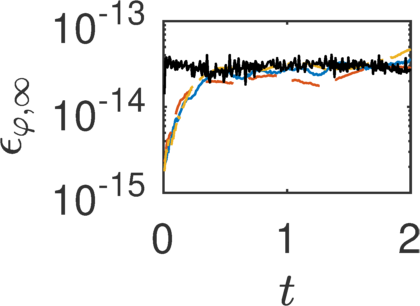}~
		\raisebox{2cm}{(b)}\includegraphics[width=3.3cm]{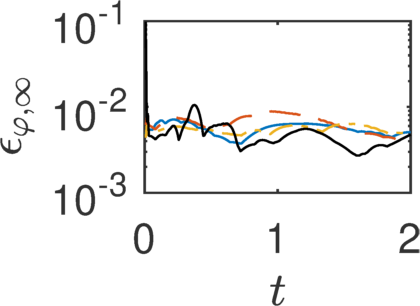}~
		\raisebox{2cm}{(c)}\includegraphics[width=3.3cm]{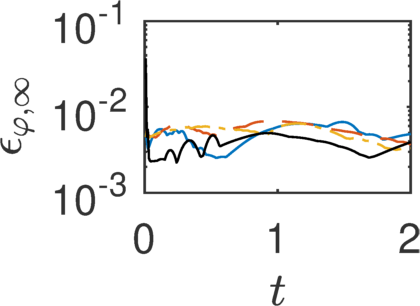}~\\
		\caption{Re-simulations errors with different velocity boundary conditions, which are (a) $\boldsymbol{u}^{*1}$, $\boldsymbol{u}^{*2}$ and $\boldsymbol{u}^{*3}$ in the corresponding directions, (b) $\boldsymbol{u}$ in all directions and (c) $\boldsymbol{u}^{*}=\boldsymbol{u}^{*3}$ in all directions. In all plots, $\Delta t=4\times10^{-3}$. See figure \ref{fig:err_ustar_vs_time} for legend.}
		\label{fig:AB2-CN}
	\end{figure}

	We demonstrate this requirement by performing 
	the original/global simulation and
	the re-simulation using the CN scheme as described above, and compare the results with cases in which some of the specific directional requirements for   
	$\boldsymbol{u}^{*}$ are relaxed. The re-simulation errors with the correct boundary condition implementation are shown in figure \ref{fig:AB2-CN}(a). The re-simulation errors remain near $10^{-14}$ for all velocities and pressure. As comparison, the re-simulations with either $\boldsymbol{u}$ or $\boldsymbol{u}^{*}=\boldsymbol{u}^{*3}$ (the last step of the ADI) velocity boundary conditions are also performed. Both produce significant error levels, between $10^{-3}$ and $10^{-2}$ (figure \ref{fig:AB2-CN}(b,c)).
	
	\section{Analysis of dominant sources of errors}\label{sec:error_analysis}
	
	In Section \ref{sec:u_or_ustar}, the correct velocity boundary conditions for re-simulation was shown to be $\boldsymbol{u}^*$. It was shown that using $\boldsymbol{u}^*$ on the boundaries based on surface data stored at every DNS time step, and replicating the precise time advancement scheme at every time step between the original DNS and the re-simulation, yielded machine-accuracy from re-simulation. However, in practical applications of STSR, one may wish to relax some of these requirements. For example, one may wish to store the boundary values not at every time-step and use moderate sub-sampling (e.g.  snapshots of the $1024^3$ isotropic turbulence data set in JHTDB are stored only every 10 simulation steps, and temporal polynomial interpolation is used to find data between stored time steps). Or, one may wish to use a different time-advancement scheme during the initial time stepping of the re-simulation. Each of these approaches will induce some additional error and prevent the re-simulation to reach machine precision. In order to establish a clear understanding of these errors, it is useful to quantify the amplification of errors by the re-simulation procedures. 
	
	In order to lay the foundation for the subsequent discussions, we 
	intentionally add noise to the boundary condition values $\boldsymbol{u}^*$. We use zero-mean Gaussian white noise and define the contaminated boundary condition on the boundary $\Gamma$, for example for the $u$-component, as
	\begin{equation}\label{eq:noise}
	{u}^*_{\sigma} = {u}^*(1 + \sigma \, G(0,1)),
	\end{equation}
	where $\sigma$ represents the root-mean-square (r.m.s.) of the added noise as multiple of the original signal. Moreover, $G(1,0)$ is a random Gaussian variable with zero mean and unit variance. Similar noise perturbations are added to the two other components $v^*$ and $w^*$, and pressure increment $\phi$, at all time steps $n>0$. 
	
\begin{figure}
		\centering
		\raisebox{3.6cm}{(a)}\includegraphics[width=5.5cm]{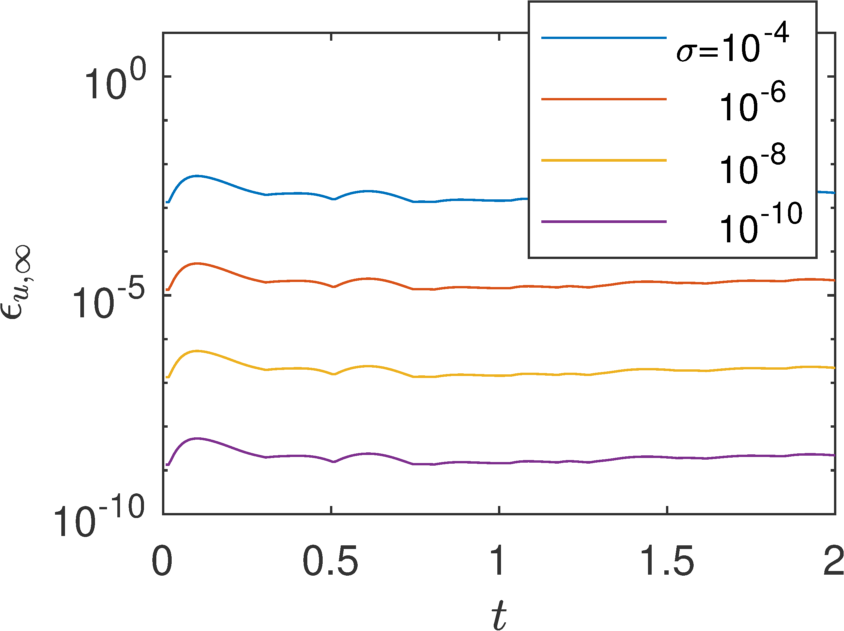}~
		\raisebox{3.6cm}{(b)}\includegraphics[width=5.5cm]{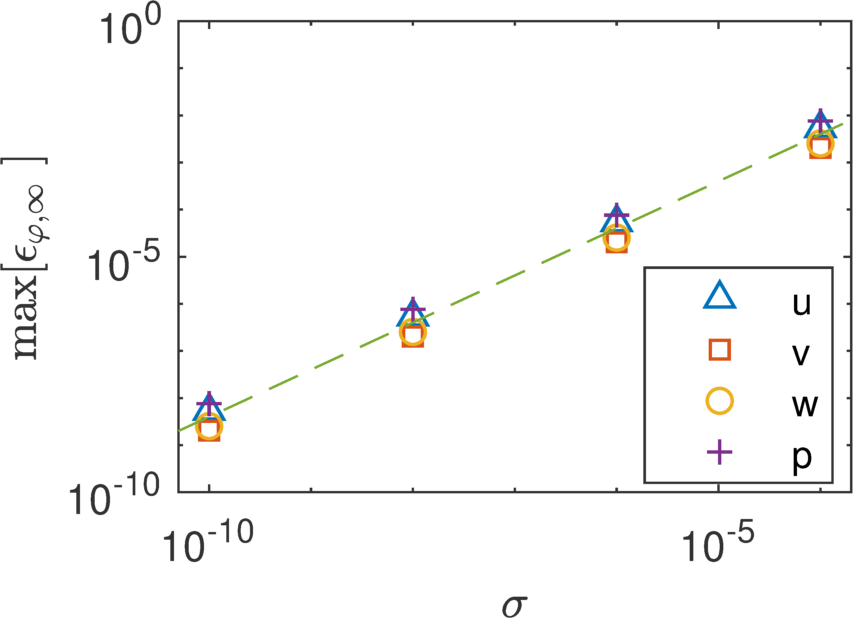}\\
		\caption{Re-simulations with different levels of noises added to the velocity boundary conditions $\boldsymbol{u}^*$. (a) Re-simulation errors $\varepsilon_{u,\infty}$ against $t$ . Only $u$ errors are plotted for clarity. It has been checked that $v$, $w$ and $p$ errors behave similarly. (b) $\max[\varepsilon_{\varphi,\infty} (t>0)]$ as function of $\sigma$. The dashed line has a slope of 1. In both plots, $\delta t=4\times10^{-3}$.}
		\label{fig:noise}
	\end{figure}

	
	Re-simulation experiments are performed for four different levels of $\sigma$ ($10^{-4}-10^{-10}$) using $\boldsymbol{u}^*_\sigma$ and $\partial \phi_\sigma/\partial \boldsymbol{n}$ as boundary conditions. The re-simulation errors $\varepsilon_{u,\infty}$ are shown in Figure \ref{fig:noise}(a) as a function of $t$ with different noise levels $\sigma$; only $u$ errors are plotted for clarity.  Although the noise levels are different, the errors are qualitatively similar at different values of $\sigma$ and only differ in magnitude. Figure \ref{fig:noise}(b) shows the scaling of $\max_t[ \varepsilon_{\varphi,\infty}]$ with $\sigma$. The results clearly show that re-simulation errors grow linearly with the magnitude of the added noise level in the boundary conditions.
	
	It should be noted that, in the above analysis, the noise is added to the boundary conditions at all time steps after the initial condition, i.e.\,$n \geqslant 1$, and the re-simulation errors are proportional to the input errors. If the noise is added at the initial condition across the entire re-simulation domain at $n=0$, similar results are obtained (not shown here).
	
	
	\subsection{Re-examination of \textbf{\textit{u}} boundary condition errors}\label{sec:err_u_as_bc}
	
	We have seen that the re-simulation errors are proportional to the input errors. We now revisit the errors discussed in Section \ref{sec:u_or_ustar}, where we first naively applied $\boldsymbol{u}$ as the velocity boundary conditions, to explain the observed errors based on the findings that errors are linearly proportional to boundary condition errors.
	
\begin{figure}
		\centering
		\raisebox{3.6cm}{(a)}\includegraphics[width=5.5cm]{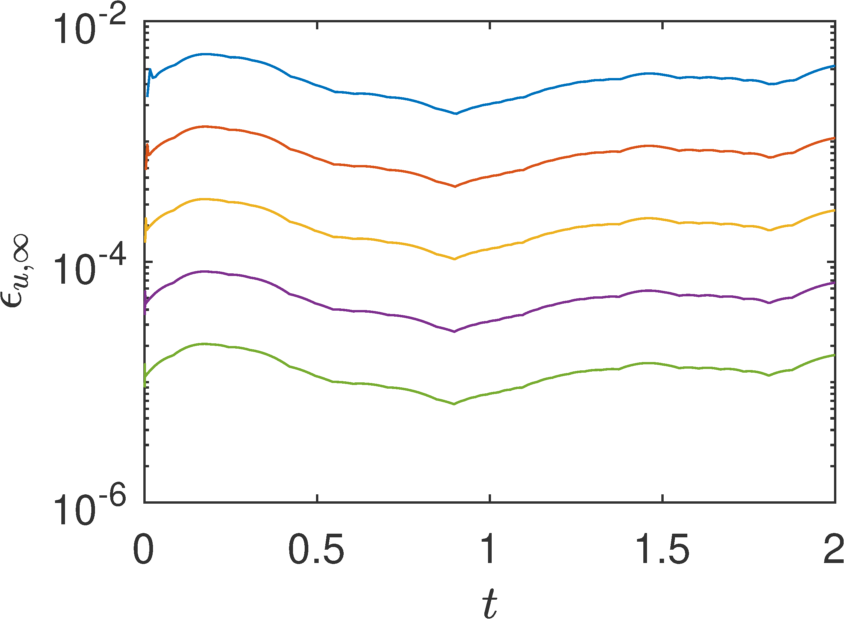}~
		\raisebox{3.6cm}{(b)}\includegraphics[width=5.5cm]{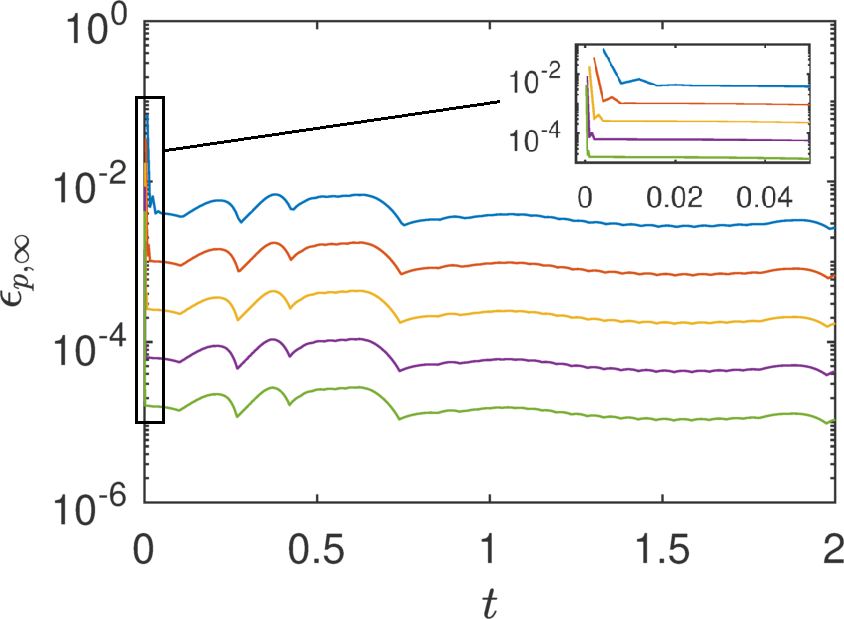}\\[7pt]
		\raisebox{3.6cm}{(c)}\includegraphics[width=5.5cm]{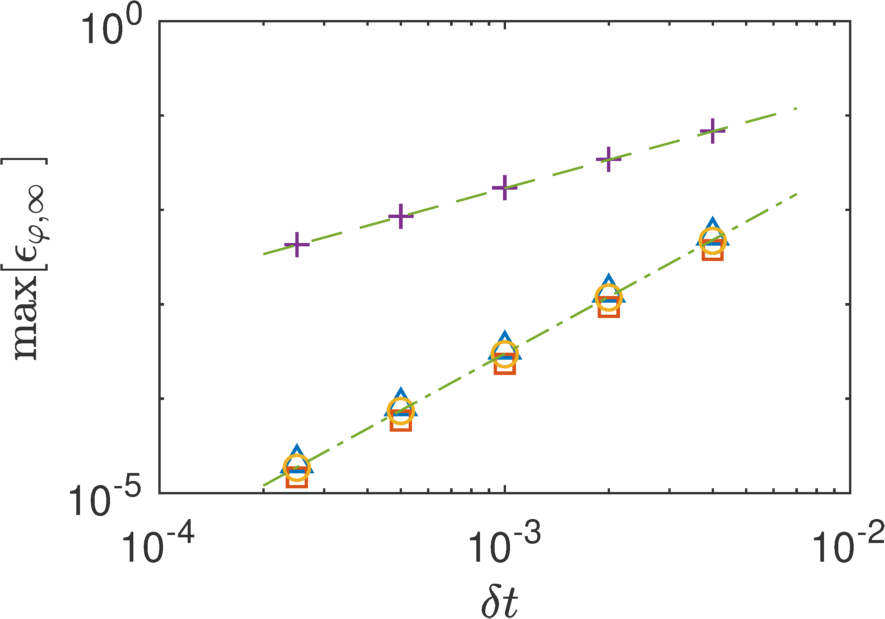}~
		\raisebox{3.6cm}{(d)}\includegraphics[width=5.5cm]{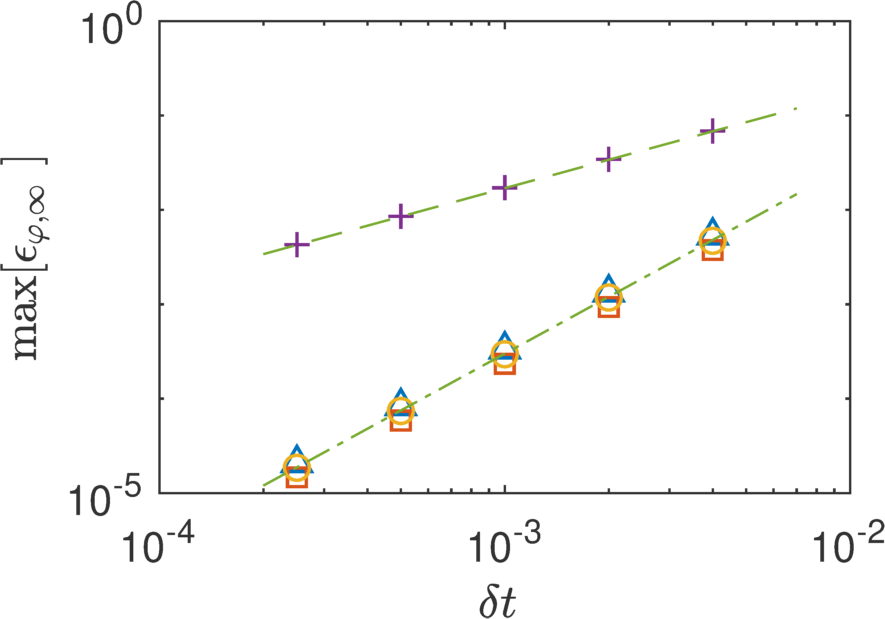}\\
		\caption{Re-simulations with $\boldsymbol{u}$ as the velocity boundary conditions using different time steps. (a) $u$ errors $\varepsilon_{u,\infty}$ as function of time, $t$ . (b) Pressure  errors $\varepsilon_{p,\infty}$ as function of time $t$. The insert is a zoom near $t=0$. In (a) and (b), lines from top to bottom represent simulations with $\delta t=4\times10^{-3}$, $2\times10^{-3}$, $1\times10^{-3}$, $5\times10^{-4}$ and $2.5\times10^{-4}$ respectively. (c) $\max[\varepsilon_{\varphi,\infty} (n\geqslant 1)]$ as function of $\delta t$. (d) $\max[\varepsilon_{\varphi,\infty} (n>1)]$ as function of  $\delta t$. In (c) and (d), the dashed line has a slope of 1 and the dashed-dotted line has a slope of 2.
		}
		\label{fig:u_as_bc}
	\end{figure}

	From equation (\ref{eq:correction}), one can easily show that the difference between $\boldsymbol{u}^{(n)}$ and $\boldsymbol{u}^*$ is second order in time, 
	\begin{equation}\label{eq:u_and_ustar}
	\boldsymbol{u}^{(n)}-\boldsymbol{u}^*=-\delta t \nabla{\phi^{(n)}}=-\delta t \nabla(p^{(n)}-p^{(n-1)}) \sim-(\delta t)^2 \, \nabla(\frac{\partial p}{\partial t}).
	\end{equation}
	Based on the results in figure \ref{fig:noise}, one would then expect that applying $\boldsymbol{u}$ as boundary conditions in the re-simulation would lead to second order errors in $\delta t$. This expectation was tested by performing the global and re-simulations with different values of $\delta t$ and prescribing $\boldsymbol{u}$ as the velocity boundary condition in the re-simulations.  The resulting re-simulation errors are plotted in figure \ref{fig:u_as_bc}(a,b). Same as in figure \ref{fig:noise}(a), $\varepsilon_{\varphi,\infty}$ behave qualitatively similar for different values of $\delta t$. The maximum errors, $\max_t[ \varepsilon_{\varphi,\infty}]$, are reported in figure \ref{fig:u_as_bc}(c). Surprisingly, the pressure errors are only first order in $\delta t$, while the velocity errors are second order, as expected. In addition, we find that the pressure errors recover second order accuracy at $n>1$ (figure \ref{fig:u_as_bc}(d)). In fact, figures \ref{fig:u_as_bc}(c) and (d) show that the maximum pressure errors are first order in $\delta t$ for $ {n\geqslant 1}$, but second order for ${n>1}$. This observation suggests that the pressure errors are of first order at $n=1$ but second order afterwards. The insert of figure \ref{fig:u_as_bc}(b) shows the pressure errors near $n=0$.
	
	A brief explanation follows:
	assume the initial field of the re-simulation matches the original global computation.
	In the first time step, if $\boldsymbol{u}^{(1)}_{os}$ is used as the velocity boundary condition, i.e., $\boldsymbol{u}^*_{\Gamma} = \boldsymbol{u}^{(1)}_{os}$, sub-domain now contain $\mathcal{O}(\delta t^2)$ errors at the boundaries,
	\begin{equation}
	\epsilon(\boldsymbol{u}^*)=
	\begin{cases}
	\boldsymbol{u}^{(1)}-\boldsymbol{u}^{*}=(\delta t)^2\frac{\partial}{\partial x}(\frac{\partial p }{\partial t})|_{n=1}=\delta t^2\zeta_{n=1} & ~ \text{on the boundaries}\\
	0  & ~ \text{inside the sub-domain}
	\end{cases},
	\end{equation}
	where $\zeta=\frac{\partial}{\partial x}(\frac{\partial p }{\partial t})$.
	From the right hand side of equation (\ref{eq:poisson}) and figure \ref{fig:u_ana}, the source term of the Poisson equation will therefore have $\mathcal{O}(\delta t)$ errors due to the errors at the sub-domain boundaries, 
	\begin{equation}\label{eq:error_p}
	DG\epsilon(\phi^{(1)})=\frac{D \epsilon(\boldsymbol{u}^*)}{\delta t} \\
	=\begin{cases}
	{\delta t^2\zeta_{n=1}}/{h\delta t}=\delta t\zeta_{n=1}/h      & ~ \text{on the boundaries}\\
	0  & ~ \text{inside the sub-domain}
	\end{cases}.
	\end{equation}
	Even though the non-zero source terms only exist at the boundary nodes in equation (\ref{eq:error_p}), the errors in $p$ contaminate the entire sub-domain due to the ellipticity of the Poisson operator.  
	Thus $\phi$ errors, as well as $p$ errors, are $h\delta t\zeta_{n=1}=\mathcal{O}(\delta t)$ at $n=1$.
	It is important to note here that $\epsilon(\phi^{(1)})$ is linearly distributed in the sub-domain (can be verified analytically to be a solution of equation (\ref{eq:error_p})).
	As a result, the gradient of $G\epsilon(\phi^{(1)})$ is uniform in the correction step, leading to a uniform $\delta t^2\zeta_{n=1}$ error in the velocity within the sub-domain:
	\begin{equation}
	\epsilon(\boldsymbol{u}^1)=\epsilon(\boldsymbol{u}^*)-\delta t G \epsilon(\phi^1) =\delta t^2\zeta_{n=1}=\mathcal{O}(\delta t^2).
	\end{equation}
	
	At the second time step $n=2$, $\boldsymbol{u}^*$ have uniform $\mathcal{O}(\delta t^2)$ errors both inside the sub-domain and on the boundaries: the errors inside the sub-domain, $\delta t^2\zeta_{n=1}$, come from $\boldsymbol{u}^{(1)}$ (see above), while the errors on the boundaries, $\delta t^2\zeta_{n=2}$, come from the new velocity boundary conditions. The leading $\mathcal{O}(\delta t^2)$ errors of $\boldsymbol{u}^*$ are cancelled out during the calculation of the divergence of $\boldsymbol{u}^*$, 
	\begin{equation}
	D \epsilon(\boldsymbol{u}^*) \\
	=\begin{cases}
	\delta t^2\zeta_{n=2}-\delta t^2\zeta_{n=1}=\delta t^3\frac{\partial \zeta}{\partial t}|_{n=1}      & ~ \text{on the boundaries}\\
	\delta t^2\zeta_{n=1}-\delta t^2\zeta_{n=1}=\mathcal{O}(\delta t^3)  & ~ \text{inside the sub-domain}
	\end{cases},
	\end{equation}
	leading to second order errors in the source term of the Poisson equation, also in the pressure field at $n=2$. In addition, the velocity errors remain at second order, 
	\begin{equation}
	\epsilon(\boldsymbol{u}^{(2)})=\epsilon(\boldsymbol{u}^*)-\delta t G \epsilon(\phi^{(2)}) = \mathcal{O}(\delta t^2)-\delta t \mathcal{O}(\delta t^2)=\mathcal{O}(\delta t^2).
	\end{equation}
	
	The preceding analysis thus demonstrates that the observed errors when using $\boldsymbol{u}$ instead of $\boldsymbol{u}^*$ as boundary conditions for re-simulation scale in expected ways with the size of time-step. If one wanted to use $\boldsymbol{u}$ instead of $\boldsymbol{u}^*$ for resimulation, however, the required time steps would be too small to be practical for purposes of the STSR.

	\subsection{Errors from mismatch in temporal discretization}\label{sec:err_time_scheme}
	
	The above results all assumed that the re-simulation starts from an Euler scheme, same as the original computation which at $n=0$ also began using an Euler step.
	This ensures that the re-simulation could calculate the intermediate velocity inside the sub-domain correctly as seen in figure \ref{fig:u_ana}, and reproduce the original simulation data precisely, when using the $\boldsymbol{u}^*_{os}$ boundary conditions.
	
	However, in  applications of STSR, the re-simulation will typically start at any of the stored original simulation time steps, i.e. when $n$ equals any integer multiple of $M_t \delta t$. Recall that the original simulation used AB2 time-stepping at those times, not Euler.
	As a result, for the re-simulation to reproduce the original computation, it must adopt an AB2 scheme from its start.  However, this requirement can only be met if two consecutive time steps are stored to be used as initial condition.  Otherwise, with a single field, the re-simulation must adopt a first Euler step and will therefore deviate from the original AB2-based computation.
	
	In order to demonstrate the errors incurred by an initial Euler step, we perform the following experiment: The data on the entire domain is stored at $t=1$, meaning the initial condition for the re-simulation is now $\boldsymbol{u}_{os}$ and $p_{os}$ at $t=1$.
	The re-simulation starts there
	with a single Euler scheme and then continues with AB2. 
	
\begin{figure}
		\centering
		\raisebox{3.6cm}{(a)}\includegraphics[width=5.5cm]{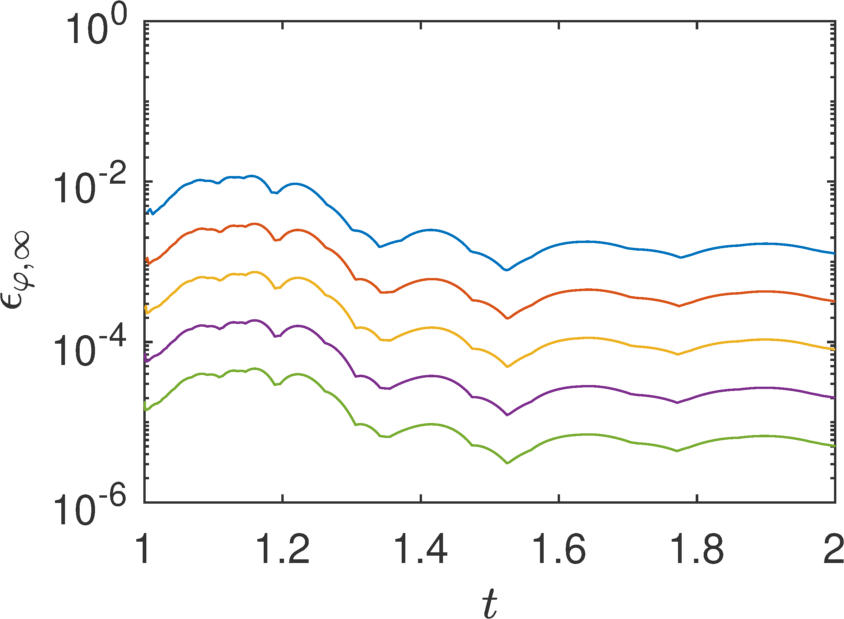}~
		\raisebox{3.6cm}{(b)}\includegraphics[width=5.5cm]{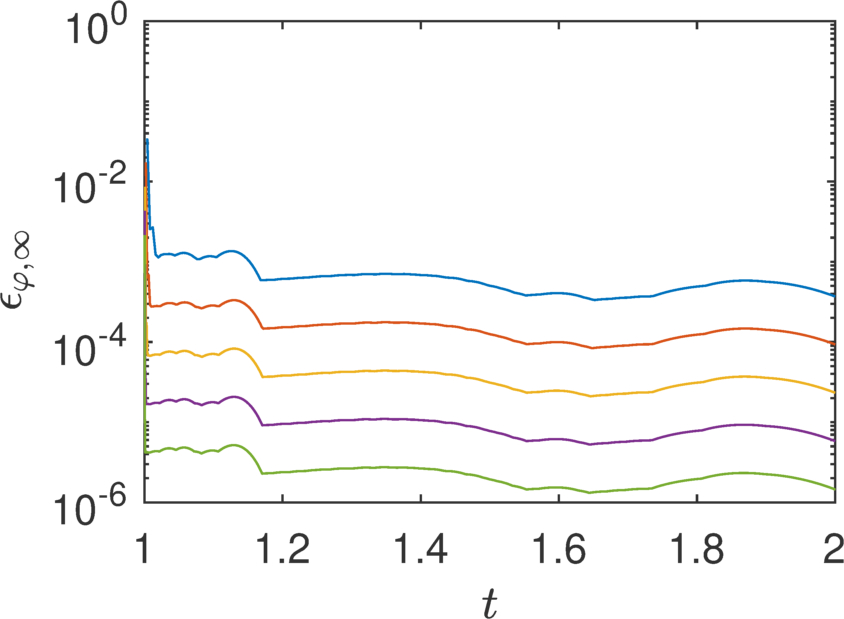}\\[7pt]
		\raisebox{3.6cm}{(c)}\includegraphics[width=5.5cm]{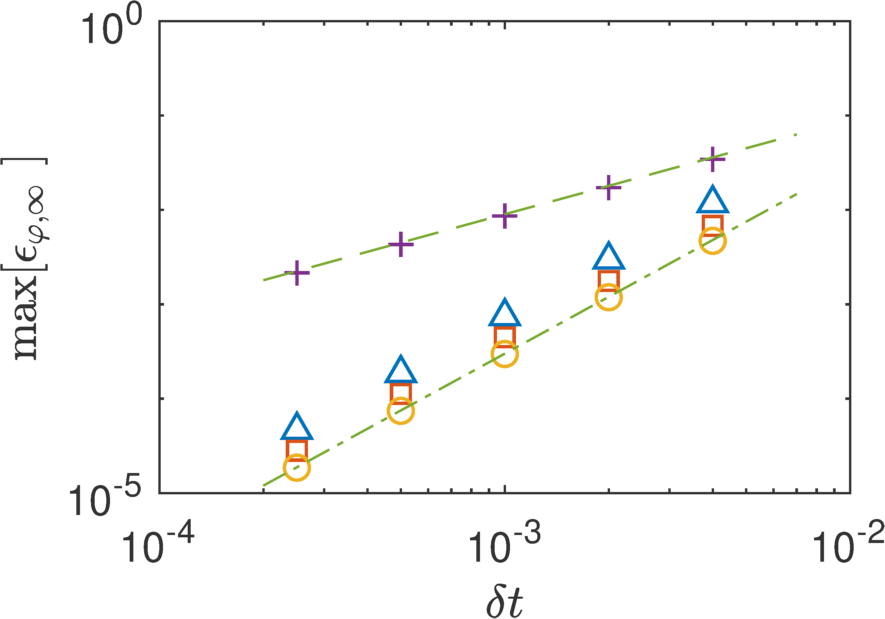}~
		\raisebox{3.6cm}{(d)}\includegraphics[width=5.5cm]{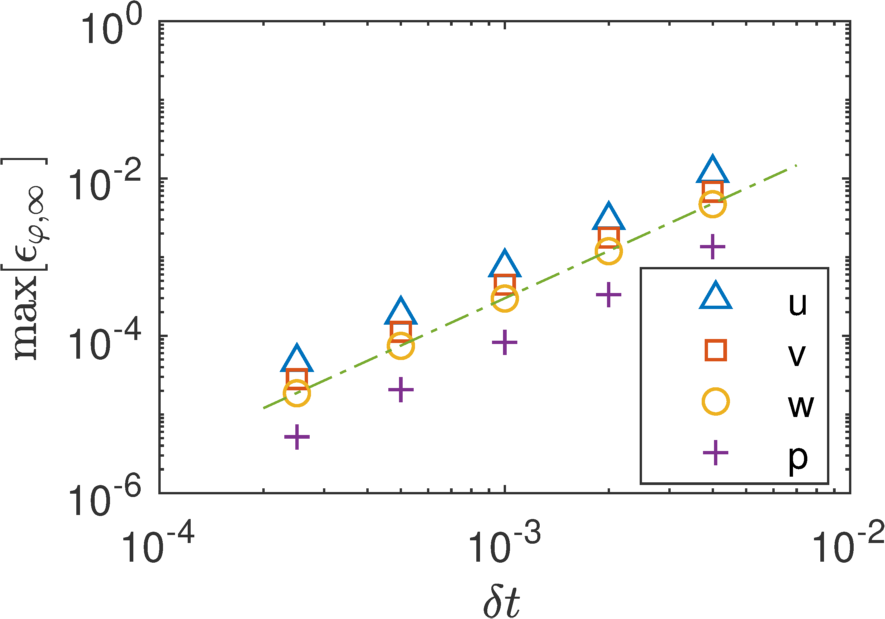}\\
		\caption{Re-simulation error evolution when using an Euler scheme at the first time step and then continuing with AB2 ($1\le t\leq 2$). The original simulation  used the AB2 scheme. (a) $u$ error $\varepsilon_{u,\infty}$ against $t$. (b) $p$ error $\varepsilon_{p,\infty}$ against $t$. In (a) and (b), lines from above to bottom represent simulations with $\delta t=\{4, 2, 1, 0.5, 0.25\}\times10^{-3}$ respectively. (c) $\varepsilon_{\varphi,\infty}$ against $\delta t$ at the first time step. (d) $\max[\varepsilon_{\varphi,\infty}]$ against $\delta t$ after the first time step. In (c) and (d), the dashed line has a slope of 1 and the dashed-dotted line has a slope of 2.
		}
		\label{fig:Euler_first}
	\end{figure}

	At the first time step after the initial condition, the Euler scheme will introduce local truncation errors of $\mathcal{O}(\delta t^2)$ into the re-simulation. The re-simulation errors are shown in figure \ref{fig:Euler_first}. Similar to the case which uses $\boldsymbol{u}$ as the velocity boundary condition (Section \ref{sec:err_u_as_bc}), the $p$ errors are first order in $\delta t$ at the first time step, but second order afterwards. On the other hand, velocity errors are always second order.
	
\begin{figure}
		\centering
		\raisebox{3.6cm}{(a)}\includegraphics[width=5.5cm]{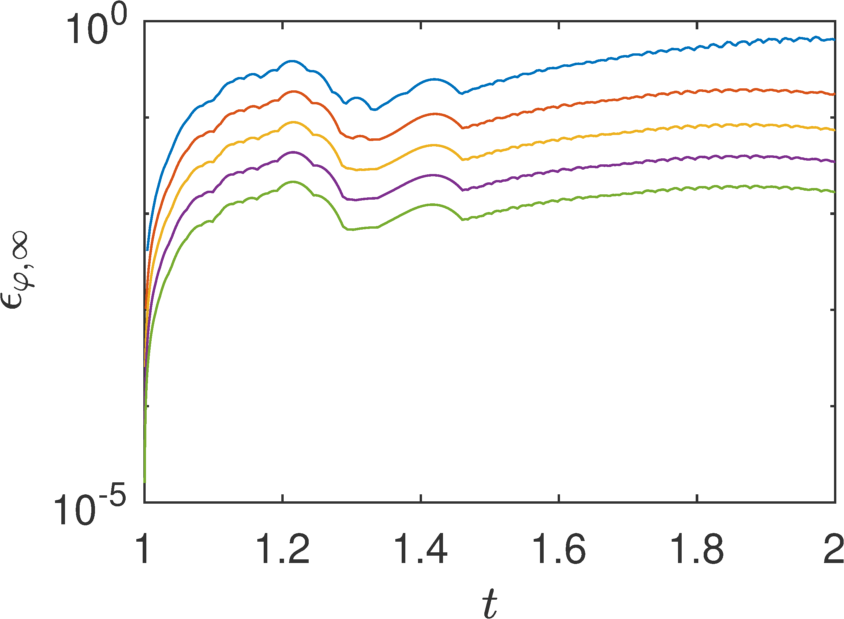}~
		\raisebox{3.6cm}{(b)}\includegraphics[width=5.5cm]{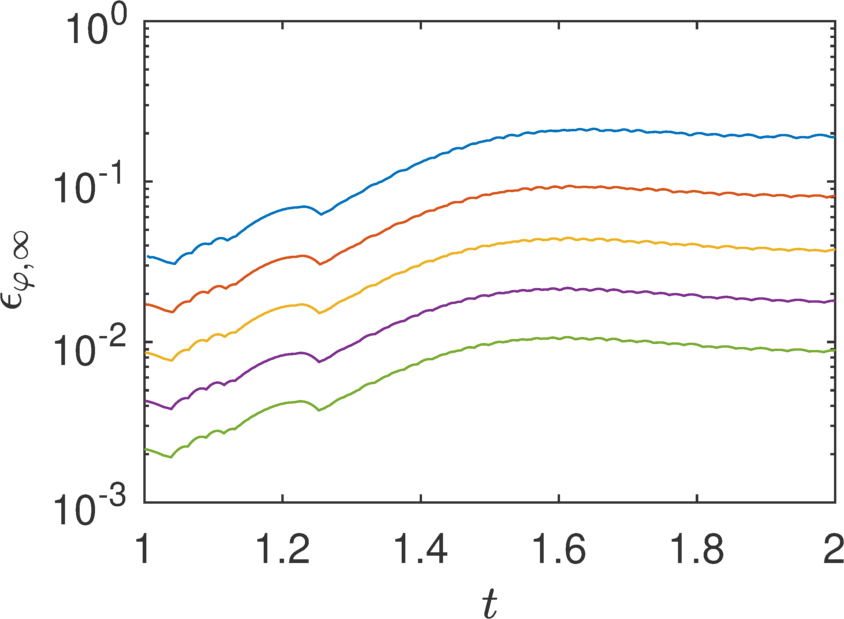}\\[7pt]
		\raisebox{3.6cm}{(c)}\includegraphics[width=5.5cm]{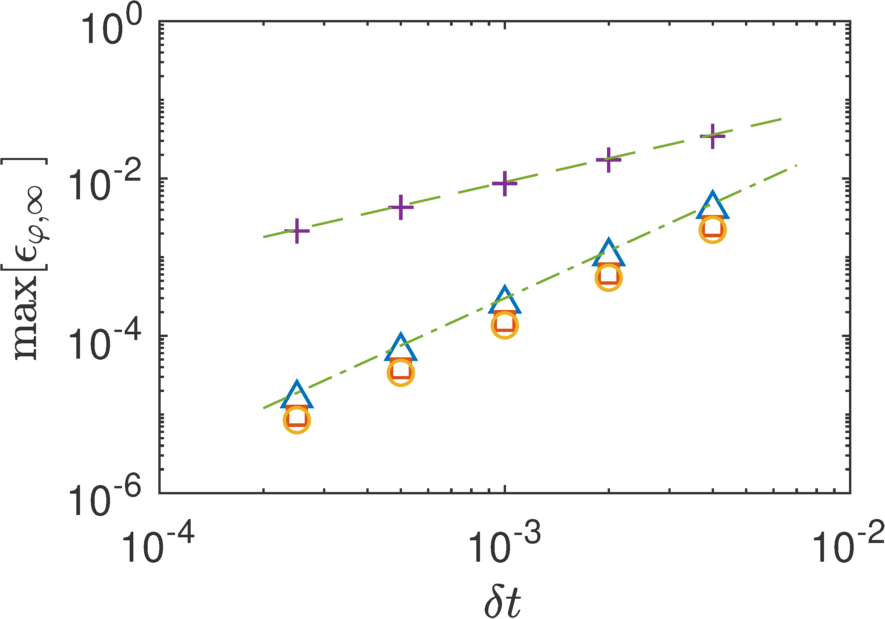}~
		\raisebox{3.6cm}{(d)}\includegraphics[width=5.5cm]{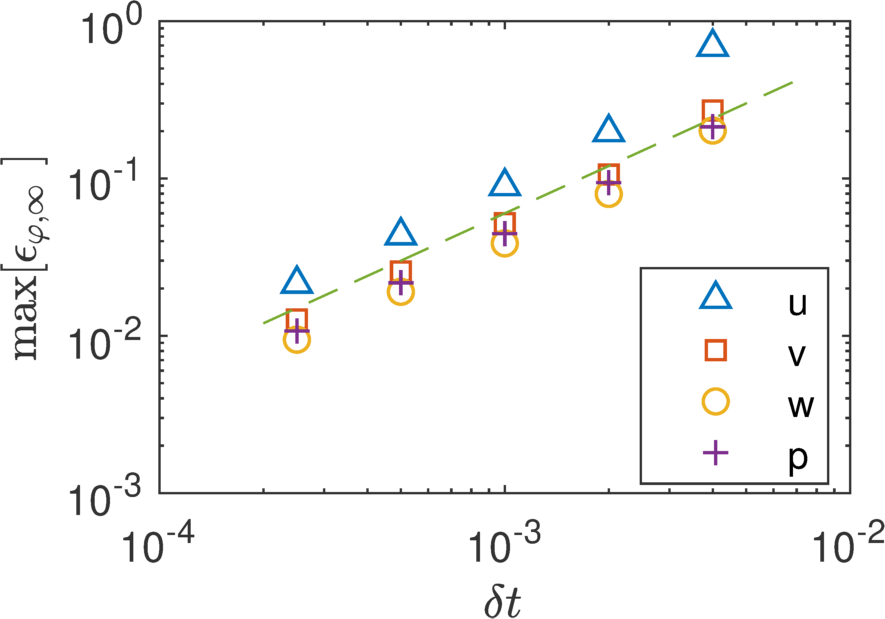}\\
		\caption{Re-simulations using and Euler time advancmeent throughout ($1\leq t\leq2$). The original simulation always uses the AB2 scheme. (a) $u$ error, $\varepsilon_{u,\infty}$, against $t$. (b) $p$ error, $\varepsilon_{p,\infty}$, against $t$. In (a) and (b), lines from above to bottom represent simulations with $\delta t=\{4, 2, 1, 0.5, 0.25\}\times10^{-3}$ respectively. (c) $\varepsilon_{\varphi,\infty}$ against $\delta t$ at the first time step. (d) $\max[\varepsilon_{\varphi,\infty}]$ against $\delta t$ after the first time step. In (c) and (d), the dashed line has a slope of 1 and the dashed-dotted line has a slope of 2.
		}
		\label{fig:Euler_always}
	\end{figure}

	In addition, we considered another case to explore errors incurred if the time stepping scheme used in the re-simulation is always different from that in the original one. We performed re-simulation with Euler scheme from $t=1$ and for all time steps, rather than for the first step only. In this case, the Euler scheme has global errors of $\mathcal{O}(\delta t)$ compared to AB2. The errors are shown in figure \ref{fig:Euler_always}.
	The $\boldsymbol{u}$ errors increase over $t$. This is due to the cumulative effect of the local truncation errors committed in each step from the Euler scheme. As a result, the velocity errors grow from second order to first order (see figure \ref{fig:Euler_always}(c-d)).
	On the other hand, the $p$ errors are already first order at the first time step, and retain that scaling, consistent with Euler's global truncation errors $\mathcal{O}(\delta t)$.
	
	%
	%

	\subsection{Errors from temporal sub-stepping}\label{sec:sub_time_step}
	
	In the previous section, it was shown that the re-simulation has $\mathcal{O}(\delta t^2)$ errors if started with an Euler scheme at an arbitrary time. These errors are too large for reproducing a DNS database using realistic values of $\delta t$. For example, when $\delta t=4\times 10^{-3}$, even if we discard the results at the first time step, the relative errors between the original and re-simulation are approximately $10^{-3}$\,--\,$10^{-2}$ in subsequent time steps.
	Using an initial Euler step in the re-simulation compared to AB2 in the original computation results in an initial error that persists in time\textemdash consistent to the behaviour when artificial errors were included in the initial conditions. 
	Although one could store an extra snapshot so that the re-simulation starts with AB2 and obtain error-free data, this approach would appreciably increase the storage requirements.
	
	Rather than store two time steps, we examine a different approach that does not increase the required storage but only increases CPU cost during re-simulation:  temporal sub-stepping.
	This idea aims to minimize the error between the original single AB2 step and many smaller steps the first of which is Euler followed by AB2. 
	
\begin{figure}
		\centering
		\includegraphics[width=\linewidth]{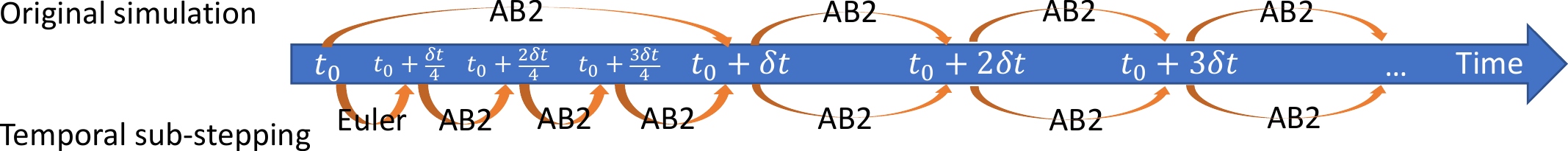}
		\caption{Schematic of temporal sub-stepping with four sub-time steps.
		}
		\label{fig:sub_tstep}
	\end{figure}

	Consider integration from $t$ to $t+\delta t$. The analytic integration could be approximated by an AB2 scheme or an Euler scheme both with a time-step size $\delta t$. We have already seen in the previous section that the differences between AB2 and Euler schemes lead to re-simulation errors. Usually, an AB2 scheme produces smaller errors than Euler compared with analytic (true) values. On the other hand, the time step from $t$ to $t+\delta t$ could also be divided into, say, $k$ sub-time steps: the size of each sub-time step is thus $\delta t/k$ (see figure \ref{fig:sub_tstep} for an example with $k=4$). Integration from $t$ to $t+\delta t$ would then be computed using Euler in the first sub-time step, then AB2 in the remaining $(k-1)$ sub-time steps. The numerical integration results will approach the true value with increasing number of sub-steps $k$. The single full-time-step Euler integration is the special case with $k=1$. Thus, one could expect that the errors between the single full-time-step AB2 integration and the integration with temporal sub-stepping would decrease first, then increase, and finally reach an asymptotic value as the number of time sub-steps $k$ increases: the asymptotic value is the errors of the AB2 scheme itself. Ideally, there will be a $k$ with which the re-simulation errors are minimized, even though this optimized $k$, if it exists, would be different from one simulation to another.
	
	Beyond $t+\delta t$, the re-simulation can proceed with AB2 using the original time step $\delta t$.  For example, the solution at $t+2\delta t$ can be computed from information at $t$ and $t+\delta t$; similarly the solution at $t+3\delta t$ can use the information at $t+\delta t$ and $t+2\delta t$ and so on. 
	
	The boundary conditions on $\Gamma$ at the sub-time steps can be approximated from temporal interpolation of $\boldsymbol{u}^*_{os}$ from the original simulation data. For instance, in the example below, the boundary conditions between $t$ and $t+\delta t$ are obtained by applying piecewise cubic Hermite interpolating polynomial (PCHIP) on stored boundary conditions (plane data) at $t-\delta t$, $t$, $t+\delta t$ and $t+2\delta t$. 
	
	For demonstration, we perform a re-simulation of the original computation with $\delta t=4\times 10^{-3}$, starting from $t=1$ and advancing the simulation until $t=2$. Re-simulations with different numbers of temporal sub-steps $k$, as well as the original AB2 scheme, are compared. Just a reminder, $k=1$ is equivalent to the re-simulation performing the entire first step with Euler scheme. In this example, the results from a re-simulation with $k=1000$ sub-time steps are used as the reference data to approximate the ``true, exact'' values which are unknown. We discard the first few $\delta t$ to avoid including the pressure jump as seen in the previous examples. 
	
	\begin{figure}
		\centering
		\raisebox{5cm}{(a)}\includegraphics[width=5.5cm]{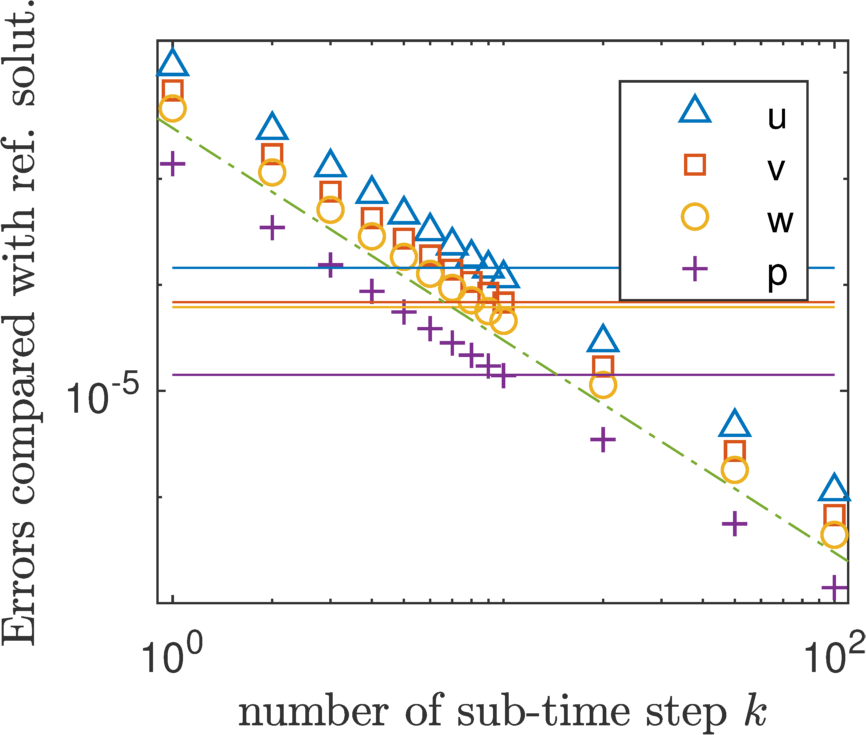}~
		\raisebox{5cm}{(b)}\includegraphics[width=5.5cm]{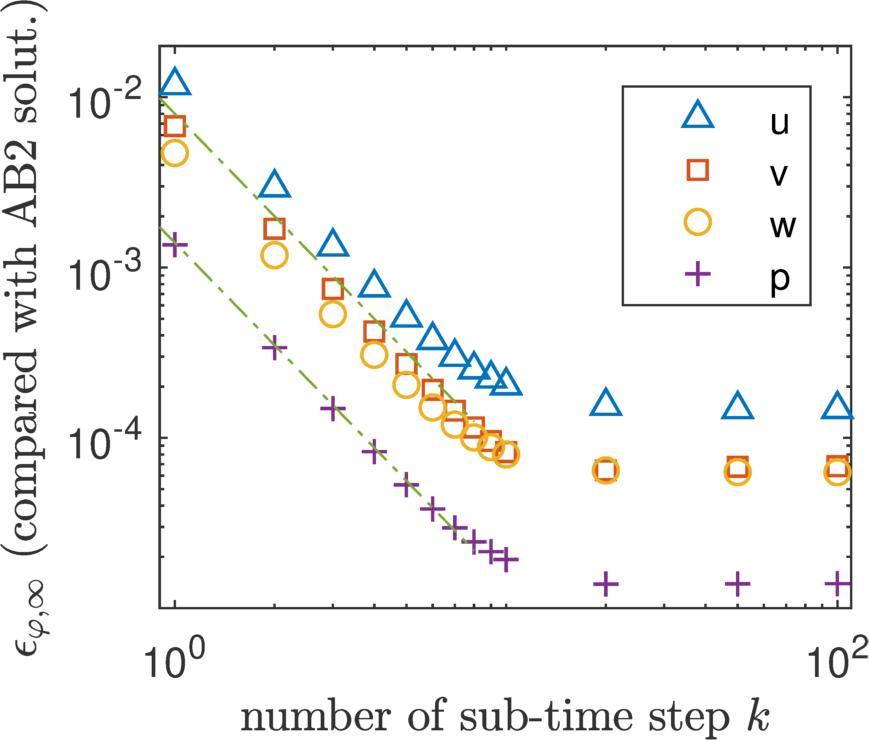}\\[7pt]
		\caption{(a) The $L^\infty$ relative errors compared with the reference re-simulation ($k=1000$). The symbols represent the re-simulations with sub-time steps, while the lines represent the original simulation with the AB2 scheme. The colours of the horizontal lines represent the same variables as the symbols.
			(b) Re-simulation errors compared with the original simulation data, $\varepsilon_{\varphi,\infty}$.
			In both plots, the dash-dot line
			has a slope of 2.}
		\label{fig:sub_tstep_res}
	\end{figure}

	Figure \ref{fig:sub_tstep_res}(a) shows the maximum relative errors compared with the reference data for $1<t<2$. The symbols denote the errors between the re-simulation and the reference data, which decrease as $k$ increases. In fact, the errors are proportional to $k^{-2}$, or the square of the size of the time sub-step $(\delta t/k)^2$, since the temporal scheme is AB2 in the re-simulation. The horizontal lines represent the errors between the original AB2 simulation and the reference data. The errors of the AB2 scheme itself are about $10^{-5}$\,--\,$10^{-4}$. Also from figure \ref{fig:sub_tstep_res}(a), it is clear that the errors between the re-simulations (symbols) and the original DNS (lines) decrease and then increase as $k$ increases. However, it should be noted that the differences between the symbols and lines do not equal to the actual errors between the re-simulations and the original DNS, $\varepsilon_{\varphi,\infty}$.

	The re-simulation errors $\varepsilon_{\varphi,\infty}$, shown in figure \ref{fig:sub_tstep_res}(b), decrease at a rate of second order in $k$ before about $k=6$, and then become nearly constant. Although an optimal $k$ is not observed, the drop of the errors is about two orders of magnitude in the current example. The asymptotic values of $\varepsilon_{\varphi,\infty}$ are also the AB2 errors shown in figure \ref{fig:sub_tstep_res}(a). This example shows that the re-simulation errors could decrease by two orders of magnitude with only 10 additional time sub-steps within the first $\delta t$ from the initial condition, and the minimum errors are bound by those of the AB2 integration scheme in the original simulation.
	
	
	
	\subsection{Temporal sub-sampling for the boundary conditions}\label{sec:temporal_subsampling_for_bc}
	
	In all previous examples, the re-simulations adopted boundary conditions data that were stored at every time step during the original DNS . This may not be necessary or feasible. As mentioned before, the snapshots of the $1024^3$ isotropic turbulence data set in JHTDB are stored only every 10 simulation steps. When data is queried between the two stored time steps, they are obtained with temporal interpolation and the errors are approximately $10^{-6}$ (we could not determine whether the interpolation errors are lower than $10^{-6}$, because the data on JHTDB are stored in single precision). Here we examine the impact of temporal interpolation of temporally sub-sampled boundary data  for re-simulation.  
	
\begin{figure}
		\centering
		\raisebox{3.6cm}{(a)}\includegraphics[width=5.5cm]{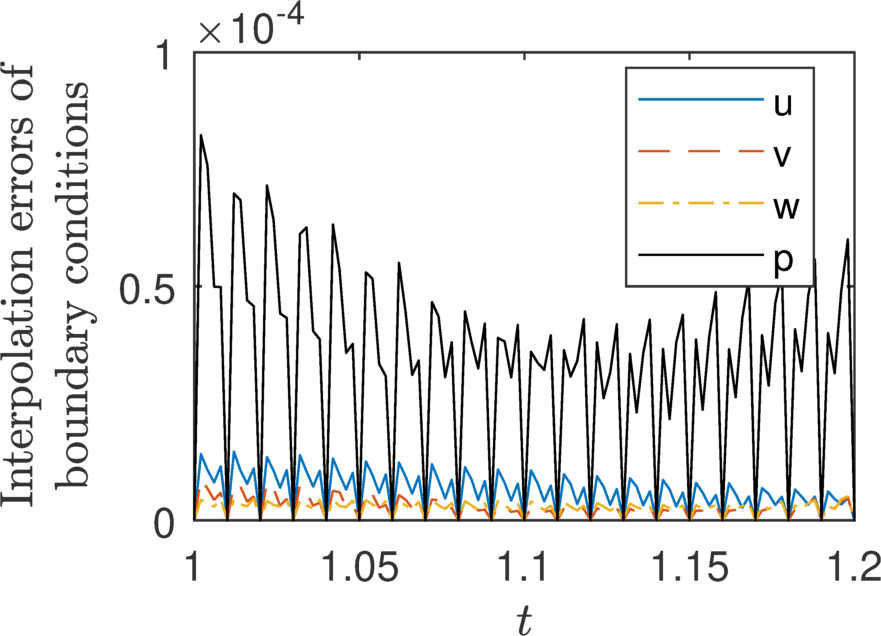}~
		\raisebox{3.6cm}{(b)}\includegraphics[width=5.5cm]{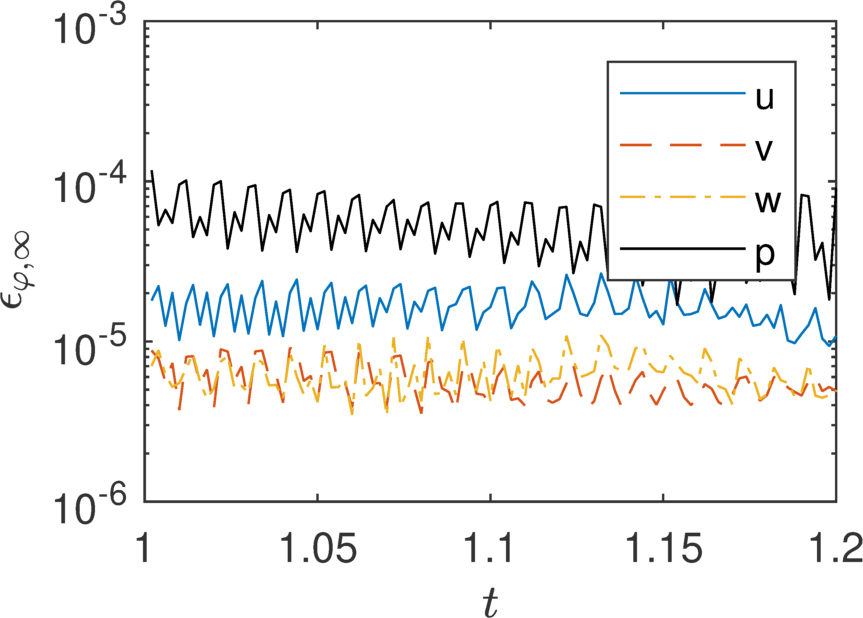}\\[7pt]
		\caption{(a) The interpolation errors of boundary conditions. (b) $\epsilon_{\varphi,\infty}$ with interpolated boundary conditions. The re-simulation is from $t=1$ to 2, but only $t=[1,1.2]$ is plotted here to more clearly display the oscillations of the errors. The re-simulation starts with the AB2 scheme using an extra snapshot provided. The time step size is $\delta t=2\times 10^{-3}$.}
		\label{fig:BC_interp}
	\end{figure}

	We have seen that the re-simulation errors are proportional to the errors in the boundary conditions. 
	Thus, if the boundary conditions are stored every few ($M_{t,bc}$) time steps and temporal interpolation is used during re-simulation, the errors in the re-simulation will be directly proportional to the interpolation errors.
	Figure \ref{fig:BC_interp} shows an example: the time step of the simulation is $\delta t=2\times 10^{-3}$. The boundary data are stored at every $M_{t,bc}=5$ time steps, actually close  to the  time step requirement based on CFL (based on maximum velocity) equaling to unity. Cubic spline interpolation with 
	three 
	points  before and after the query point is used for temporal interpolation.
	The $L^\infty$ relative errors of the interpolated   boundary condition fields on the $\Gamma$ planes are shown in figure \ref{fig:BC_interp}(a). The oscillations of the errors are apparent, vanishing at each of the $5\delta t$ time instants in which boundary data are known exactly.
	The re-simulation starts at $t=1$ using the AB2 scheme with an extra snapshot provided, and runs until $t=2$.  As a result, no other errors are introduced in the re-simulation, except those due to the temporal interpolation of the boundary conditions.  The maximum interpolation errors over time for $\{u, v, w, p\}$ are \{1.47, 1.54, 1.64, 9.66\}$\times 10^{-5}$ (figure \ref{fig:BC_interp}(a)).
	The re-simulation errors $\epsilon_{\varphi,\infty}$ (figure \ref{fig:BC_interp}(b)) for $\{u, v, w, p\}$ are \{2.66, 2.08, 2.30, 24.2\}$\times 10^{-5}$: all are only slightly higher than the interpolation errors. The oscillations of the re-simulation errors are caused by the oscillatory errors of the temporal interpolation of the boundary conditions. 
	
	
	\section{Summary: recommended choices for STSR}
	\label{sec:summary}
	
	The previous section has documented separately errors to be expected from various parameter choices for STSR. Here we now combine the various choices that may be expected in an actual implementation of STSR: we use $\boldsymbol{u}^*$ on the boundaries stored at every $M_{t,bc}=5$ DNS time steps, use $k=10$ for the initial temporal sub-sampling during the first time-step of re-simulation, use cubic polynomial temporal interpolation of the stored $\boldsymbol{u}^*$ and $p$ boundary values to interpolate to the re-simulation time-step $\delta t$, and integrate between $t=1$ and $t=2$. 
	
\begin{figure}
		\centering
		\raisebox{4.5cm}{(a)}\includegraphics[width=5cm]{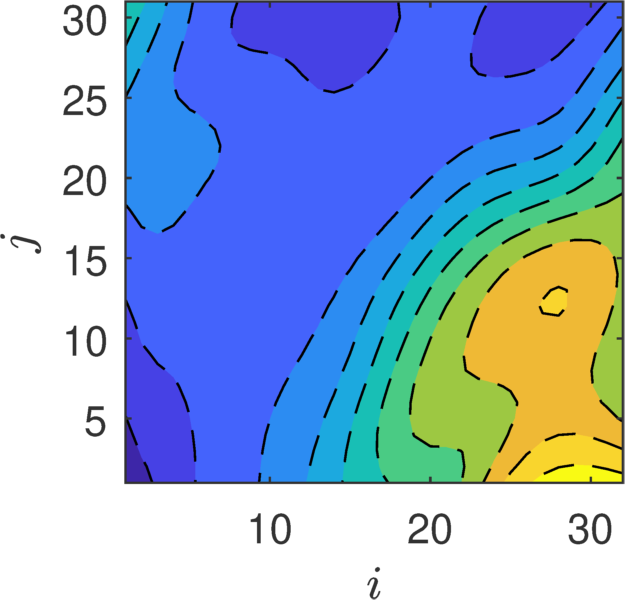}~
		\raisebox{4.5cm}{(b)}\includegraphics[width=5cm]{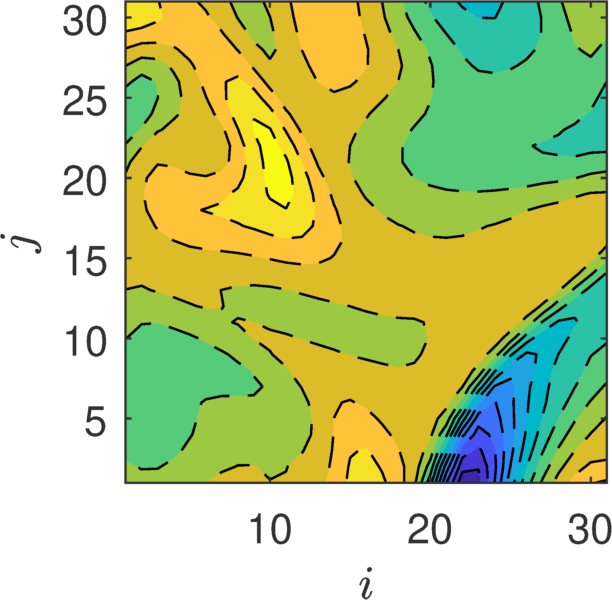}\\[7pt]
		\raisebox{4cm}{(c)}\includegraphics[width=11cm]{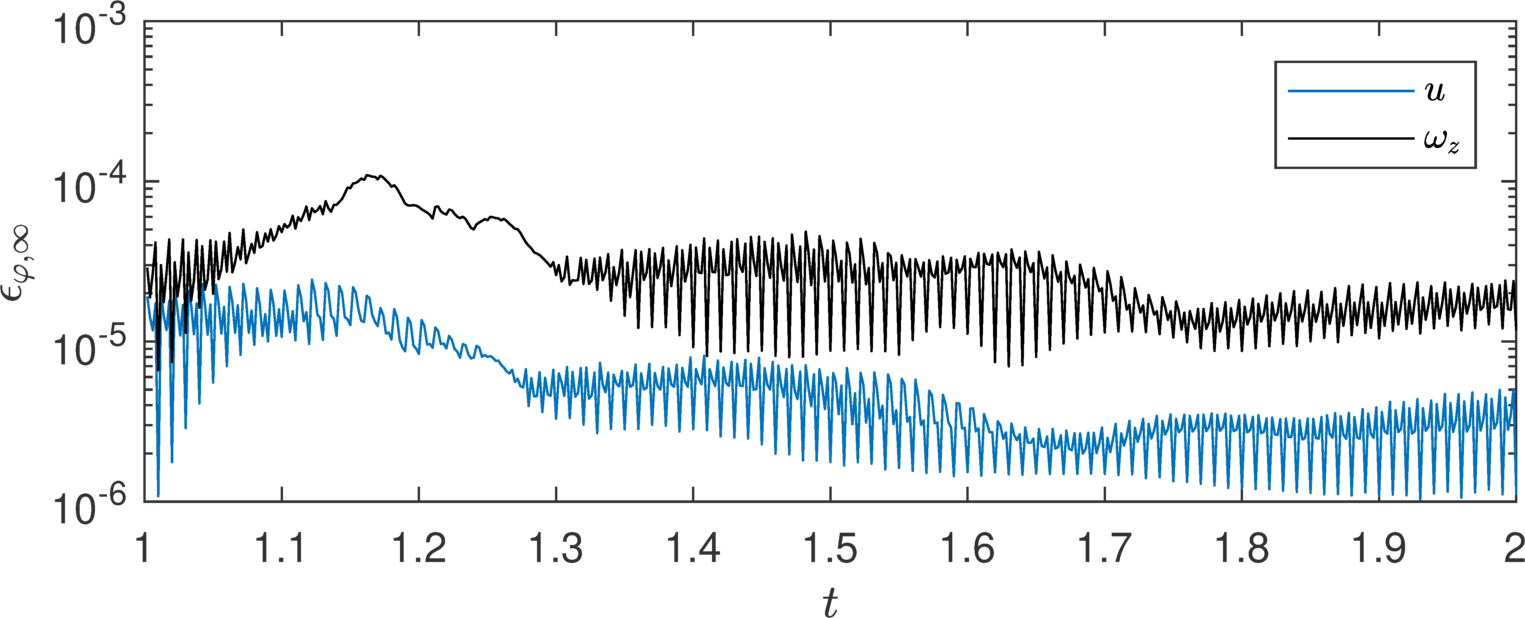}\\[7pt]
		\caption{(a) Contour plot of $u$ on a randomly selected slice. (b) Contour plot of $z$-component vorticity on a randomly selected slice. In (a) and (b), colour contours are the original simulation, while the black dash contour lines are the re-simulation. (c) $L^\infty$ errors of $u$ and $z-$ component vorticity.}
		\label{fig:combined}
	\end{figure}

	Figure \ref{fig:combined} compares two fields at $t=2$ from the re-simulation to the original simulation: (a) $u$-velocity and (b) $z-$component vorticity $\partial v/\partial x - \partial u/\partial y$ (computed using centered finite differencing). The contour lines of re-simulation fields and the original ones are on top of each other.
	
	Figure \ref{fig:combined}(c) shows the corresponding evolutions of the $L^\infty$ errors. The vorticity errors are about one order of magnitude higher than velocity errors and is about $10^{-4}$. This level of difference between re-simulation and original DNS is acceptable and falls within the desired guidelines.
	
	\section{Conclusions}\label{sec:conclusion}
	
	In the present paper, we propose an idea of data compression for numerical simulation results of incompressible fluid flow. The entire simulation domain of the original simulation is divided into multiple small sub-regions by planes. The data in the entire domain are stored, say, at every few hundred or thousand time steps, while data on the dividing planes are stored at every time step, or subsampled every few time steps. Once data at an arbitrary position and time  is needed, a re-simulation of the small cube region (sub-domain) which includes that point is performed. The data stored in the entire domain are used as the initial condition, while the planar data surrounding the sub-domain are used as the boundary conditions.
	
	It is found that if the numerical scheme in the re-simulation matches the original simulation exactly, the re-simulation will produce error-free results. On the other hand, any mismatch between the re-simulation and the original one can produce significant errors, exceeding the minimum error levels one would like to enforce for a database that contains spatially and temporally subsampled data.
	
	For example, it was found that re-simulation errors are too high when using  velocity and pressure differences (or pressure) for the boundary condition. It was shown that the correct velocity boundary conditions for the re-simulation should be the intermediate velocity after the projection step: this is because the boundaries of the sub-domain are still the internal part of the entire domain of the original simulation.
	
	Another example is that the re-simulation should use the same time integration scheme as the original simulation. This poses a challenge if only one snapshot of the initial field is provided: the re-simulation must start with an Euler scheme while the original simulation has been advanced with an AB2 scheme. The challenge can be resolved by storing an extra snapshots so that the re-simulation could start with the AB2 scheme as well, or could be improved using Euler-AB2 integration with several sub-time steps to approximate the first AB2 integration in the original simulation. We have shown the latter approach saves storage space, and can also reduce the re-simulation errors by two orders of magnitude with only 10 sub-time steps added in the first original time step.
	
	Tests using boundary data with added noise show that re-simulation errors remain linearly proportional to the errors in the boundary conditions. This observation helps explain several trends in re-simulation errors. Also, it provides a guideline about how much temporal sub-sampling of the boundary data may be used. The resulting errors in re-simulation will be proportional to the errors caused by temporal interpolation on the boundary data.
	Experiment shows the re-simulation error is 
	similar to the interpolation errors of the boundary conditions. Thus, in a real application, one could carefully control the interval of two stored plane data and achieve further compression of the simulation data.
	
	A sample application combining all of the recommended sub-sampling parameters and re-simulation strategies shows that relative maximum errors in velocity on the order of $10^{-5}$  to $10^{-4}$, which is acceptable and leads to errors of less that $0.1\%$ in  velocity gradients. 
	These levels are acceptable for applications of building numerical turbulence databases like JHTDB. 
	
	Finally, we remark that  alternative resimulation methods e.g. based on machine learning instead of Navier-Stokes could be considered. For instance, one could train an Artificial Neural Network to predict field data at desired points and time using similar types of initial and bounding surface data as used in the present method as inputs. The present results documenting errors to be expected from Navier-Stokes based re-simulation can serve as useful reference or benchmark to which to compare such alternative methodologies.
	
	\section*{Acknowledgments}
	The authors acknowledge funding from the National Science Foundation (grant \# OCE-1633124). Computations were made possible by the Maryland Advanced Research Computing Center (MARCC). Useful discussions and conversations with Profs. A. Szalay, R. Burns, G. Eyink and Dr. C. Lalescu are gratefully appreciated. 
	
	\section*{Appendix A: Fast Poisson solver for re-simulation}
	
	In this appendix, details about a spectral fast Poisson solver for equation (\ref{eq:poisson}) used in re-simulations are described. Since the re-simulation sub-domain is in general not periodic, a fast Poisson solver using discrete sine and cosine transforms \citep{Schumann1988} is implemented.  
	
	Consider a one-dimension Poisson equation, 
	\begin{equation}\label{eq:1d_poisson}
	\nabla^2 \psi=b
	\end{equation}
	on a uniform grid $x_i=i h\ (i=1,\dots,N)$, where $h=\Delta x$ is the constant grid spacing. The Poisson equation discretized with second-order central finite differences is
	\begin{equation}\label{eq:dis_pos}
	\frac{\psi_{i-1}-2\psi_i+\psi_{i+1}}{h^2}=b_i,\qquad i=1,\dots,m,
	\end{equation}
	and can be represented in Fourier space as
	\begin{equation}\label{eq:dis_pos_fourier}
	\lambda_j \hat{\psi}_j=\hat{b}_j,\qquad j=1,\dots,N,
	\end{equation}
	where $\lambda=-k'^2$ is the eigenvalue and $k'$ is the modified wavenumber. 
	Thus, the Poisson equation can be solved in three steps:
	(i) calculate $\hat{b}_j$ from the forward Fourier/sine/cosine transform of $b$; 
	(ii) find $\hat{\psi}_j=\hat{b}_j/\lambda_j$ from equation (\ref{eq:dis_pos_fourier});
	(iii) calculate $\psi$ from the inverse transform of $\hat{\psi}_j$.
	The transforms used in (i,iii) and the eigenvalues $\lambda_j$ depend on the boundary conditions and are listed in tables \ref{tab:transform} and \ref{tab:eig}. In table \ref{tab:transform}, ``DFT'' refers to the discrete Fourier transform, ``DST-II'' to type-II discrete sine transform, and ``DCT-II'' to type-II discrete cosine transform.
	For non-homogeneous boundary conditions, $b_1$ and $b_n$ can be modified in order to absorb the values at the boundaries.  
	
	{
		\setlength{\tabcolsep}{1.5em}
		\begin{table}
			\begin{center}
				\def~{\hphantom{0}}
				\begin{tabular}{lcc}
					\hline
					Boundary conditions & Forward & Backward \\[5pt]
					\hline
					Periodic ($x_0=x_m,\ x_{m+1}=x_1$) & DFT & Inverse of DFT \\[7pt]
					\begin{tabular}{@{}l@{}}Dirichlet on cell faces \\ ($x_1+x_0=0,\ x_{m+1}+x_{m}=0$)\end{tabular} & DST-II & Inverse of DST-II \\[15pt]
					\begin{tabular}{@{}l@{}}Neumann on cell faces \\ ($x_1-x_0=0,\ x_{m+1}-x_{m}=0$) \end{tabular} & DCT-II & Inverse of DCT-II \\[3pt]
					\hline
				\end{tabular}
				\caption{The transforms used in steps 1 and 3 in the fast Poisson solver.}
				\label{tab:transform}
			\end{center}
		\end{table}
		
		\begin{table}
			\begin{center}
				\def~{\hphantom{0}}
				\begin{tabular}{ll}
					\hline
					Boundary conditions & Eigenvalues \\[5pt]
					\hline
					Periodic ($x_0=x_n,\ x_{m+1}=x_1$) & $\lambda_k = -\frac{4}{h^2}\sin^2 \frac{(k-1)\pi}{m}$ \\[7pt]
					\begin{tabular}{@{}l@{}}Dirichlet on cell faces \\ ($x_1+x_0=0,\ x_{m+1}+x_{m}=0$)\end{tabular} & $\lambda_k = -\frac{4}{h^2}\sin^2 \frac{k\pi}{2m} $ \\[15pt]
					\begin{tabular}{@{}l@{}}Neumann on cell faces \\ ($x_1-x_0=0,\ x_{m+1}-x_{m}=0$) \end{tabular} & $\lambda_k = -\frac{4}{h^2}\sin^2 \frac{(k-1)\pi}{2m}$ \\[3pt]
					\hline
				\end{tabular}
				\caption{The eigenvalues used in step 2 in the fast Poisson solver.}
				\label{tab:eig}
			\end{center}
		\end{table}
	}
	
	When $\lambda_1=0$, an additional equation is required, e.g. with the periodic or Neumann boundary conditions in all directions one could simply set $\hat\psi_1=0$ leading to a zero-mean solution. It should also be noted that this algorithm gives the least square solution for the discretized Poisson equation if the compatibility condition $\sum{b_i}=0$ is not satisfied.
	
	The discrete Fourier, sine and cosine transforms are included in various libraries, including FFTW and FFTPACK. If a DST-II or DCT-II is not implemented, e.g.\,in the Intel Math Kernel Library (MKL), it can be computed via a DCT-III combined with $\mathcal{O}(2n)$ pre- and post-processing.
	
	Extension of the algorithm to 3D is straightforward: (i) calculate $\hat{b}_{j_1 j_2 j_3}$ from the forward transform of $b$;
	(ii) find $\hat{\psi}_{j_1 j_2 j_3}=\hat{b}_{j_1 j_2 j_3}/\lambda_{j_1 j_2 j_3}$, where $\lambda_{j_1 j_2 j_3}=\lambda_{j_1}+\lambda_{j_2}+\lambda_{j_3}$;
	(iii) calculate $\psi$ from the backward transform of $\hat{\psi}_{j_1 j_2 j_3}$.  
	
	If the grid is non-uniform in only one direction, e.g. in channel or boundary-layer flows, the spectral approach is adopted in all dimensions where the grid is uniform, and a tri-diagonal solver is adopted in the direction of grid stretching (see \citet[Section 6.2.1]{Moin2010} for an example). In fact, solving a tri-diagonal linear system is faster than Fourier transforms, since the former has a computational cost $\mathcal{O}(N)$, which is less than that of fast Fourier transform, $\mathcal{O}(N\log N)$.
	
	The current fast Poisson solver is faster in time and saves the memory compared with a Poisson solver implementing sparse matrix solver. Table \ref{tab:poisson_time} compares the time spent in solving the discrete Poisson equation using sparse matrix LU decomposition, FFT and DST/DCT. When the gird comprises $128^3$ points, the LU decomposition requires extensive memory and in our tests using limited resources (as one would like to use during re-simulation), it runs out of memory.
	The solution using DST/DCT requires approximately twice the time of the DFT, and only one-dimensional DST/DCT are available in the majority of numerical libraries. Nevertheless, DST/DCT outperforms the direct solver based on the sparse matrix LU decomposition, and its scalability is superior.

	\begin{table}
		\begin{center}
			\def~{\hphantom{0}}
			\begin{tabular}{ccccc}
				\hline
				Grid points & LU decomposition
				& DFT  & DST/DCT  \\[5pt]
				\hline
				$32^3$ & 0.0082 s & $<10^{-3}$ s & $<10^{-3}$ s  \\[7pt]
				$48^3$ & 0.0357 s & $<10^{-3}$ s & 0.0016 s  \\[7pt]
				$64^3$ & 0.1137 s & 0.0019 s & 0.0035 s  \\[7pt]
				$96^3$ & 0.5451 s & 0.0052 s & 0.0101 s  \\[7pt]
				$128^3$ & - & 0.0109 s & 0.0234 s  \\[7pt]
				\hline
			\end{tabular}
			\caption{Time spent in solving the discrete Poisson equation with a sparse matrix solver, FFT or DST/DCT. The timing has a resolution of $10^{-3}$ s, and is averaged over 100 runs.
				In the LU decomposition method, only the solution phase (i.e. forward and backward substitutions after the LU decomposition) is timed.
				The hardware is Intel Core i5-7500 (4 Cores, 3.4GHz) and 16GB memory.
				The code uses Intel Fortran compiler, Intel MKL and OpenMP in Windows.
				The parallelization of the sparse matrix solver and the DFT is implemented in Intel MKL, while that of DST/DCT is implemented by authors using OpenMP.
				In the $128^3$ case, the LU decomposition runs out of memory.}
			\label{tab:poisson_time}
		\end{center}
	\end{table}

	\bibliography{paper2}
	
\end{document}